\documentclass{article}

\usepackage{graphicx}
\usepackage{subfigure}
\usepackage[authoryear]{natbib}
\usepackage{amsfonts}

\title{A Bayesian approach to inferring the phylogenetic structure of communities from metagenomic data}
\author{John D. O'Brien$^{1}$, Xavier Didelot$^{2}$, Zamin Iqbal$^{3}$,\\ Lucas Amenga-Etego$^{3}$, Bartu Ahiska$^{4}$, Daniel Falush$^{5}$}

\date{~ }

\usepackage{fullpage}
\begin{document}
\maketitle
\noindent $^{1}$ Department of Mathematics, Bowdoin College, Brunswick, Maine 04011, USA 

\noindent $^{2}$ School of Public Health, Imperial College London, London W2 1PG, United Kingdom 

\noindent $^{3}$ Wellcome Trust Centre for Human Genetics, University of Oxford, Oxford OX3 7BN, United Kingdom 

\noindent $^{4}$ Department of Statistics, University of Oxford, Oxford OX1 3TG, United Kingdom

\noindent $^{5}$ Max Plank Center for Evolutionary Anthropology, Deutscher Platz 6, 04103 Leipzig, Germany

\clearpage

Running Head: Bayesian phylogenetics for community metagenomics\\


Corresponding authors: \\ 

John D. O'Brien

Bowdoin College

Department of Mathematics

8600 College Station

Brunwswick, Maine, United States  04011

Phone: +(001) 207 798 4247

Fax: +(001) 207 725 3750

Email: \emph{jobrien@bowdoin.edu}

\vspace{1cm}

Daniel Falush

Max Plank Center for Evolutionary Anthropology

Deutscher Platz 6

04103 Leipzig, Germany

Phone: +49 (0) 341 35 50 500 

Fax: +49 (0) 341 35 50 555

Email: \emph{daniel\_falush@eva.mpg.de}

\clearpage

\begin{abstract}
{\normalsize Metagenomics provides a powerful new tool set for investigating evolutionary interactions with the environment.  However, an absence of model-based statistical methods means that researchers are often not able to make full use of this complex information.  We present a Bayesian method for inferring the phylogenetic relationship among related organisms found within metagenomic samples.  Our approach exploits variation in the frequency of taxa among samples to simultaneously infer each lineage haplotype, the phylogenetic tree connecting them, and their frequency within each sample.  Applications of the algorithm to simulated data show that our method can recover a substantial fraction of the phylogenetic structure even in the presence of strong mixing among samples.  We provide examples of the method applied to data from green sulfur bacteria recovered from an Antarctic lake, plastids from mixed \emph{Plasmodium falciparum} infections, and virulent \textit{Neisseria meningitidis} samples.}{\normalsize \par}
\end{abstract}
\clearpage

\section{Introduction}
Metagenomics -- purifying and sequencing DNA from environmental samples without any culturing step -- represents an important new tool for investigating how microbes interact with, mold and adapt to their environments \citep{Allen2005,Tyson2004,Gill2006,Preidis2009}.  Metagenomics can also be applied to any situation where genetic variability exists within a sample, such as microbiomes, mixed infections, and cancer.  Many metagenomic analyses relate the overall DNA content of samples to environmental phenotypes \citep{Green2005,Kurokawa2007}.  We take up a different problem: the reconstruction of organismal composition for each sample.  Overall DNA content provides useful information on overall community function but many physiological and evolutionary processes may only be understood at the organismal level \citep{Martinez2005,Martinez2009}.

Recent improvements in sequencing technology allow the collection of large numbers ($>10^6$) of short reads of DNA sequence ($40-100$ bp) from within a sample \citep{Schmeisser2007,Bentley2008}. For notational clarity we refer to each sample as a \emph{pool}.  The simplest approach to inferring composition is in terms of the frequency of known sequences within each sample \citep{vonMering2007,Chaffron2010}. This approach typically works well for assessing variation at broad scales when individual reads can be mapped onto the nearest reference genome within the tree of life. However, at finer scales, and in particular if one is interested in the evolution taking place within the samples themselves, the structure of relationships among organisms will generally not be known in advance and so must be inferred from data.

The left-hand side of Figure 1 illustrates the evolutionary scenario that we assume underlies the data.  The phylogeny's tips correspond to individual cells and color indicates the pool of origin.  Since individual reads are typically short, and will thus contain limited phylogenetic information, it is not feasible to reconstruct a resolved tree where each read corresponds to a single taxon.  We therefore attempt to infer a simplified phylogeny in which the terminal nodes represent groups of related organisms, or \emph{lineages} (right-hand side of Figure 1).  Each lineage defines a haplotype of allele states for the single nucleotide polymorphisms (SNPs) within the data and makes up a proportion of the organisms within a pool shown by the colored bar.  As indicated by the shaded \emph{cones}, the SNP pattern of organisms within a lineage may vary, perhaps due to sequencing errors or low-frequency variation. 

One similar -- but easier -- problem is phasing in diploid organisms. In this case, the goal is to reconstruct haplotypes (i.e. the sequences of the two copies of each chromosome) given the genotypes at each diploid locus. Statistical algorithms often estimate phase using the property that particular combinations of variants are present in the population at a higher frequency than expected if the variants segregated independently \citep{Excoffier1995,Stephens2001}. In our case, we seek to estimate the underlying lineages and the phylogeny connecting them using the differing frequencies of SNP allele proportions within each pool.

We focus on extracting the phylogenetic information provided by this SNP read count variation.  Our model assumes that information comes independently from SNPs and neglects information either from multiple SNPs co-occurring on a single read or on paired-end reads.  This means that we discard potentially valuable linkage data that provides strong information about haplotype structure.  Other algorithms have be developed that specifically seek to utilize this data \citep{Greenspan2004} and we will briefly detail the prospects for improvements that exploit both variation between pools and the information from linked SNPs in the discussion.

\section{Data and Methods}

\subsection{Model}
Our model is primarily a phylogenetic one, and so borrows a lot of its structure from established methods \citep{Mau1999,Felsenstein2004,Drummond2005}.  However, our data are distinct from standard phylogenetic contexts since individual metagenomic reads cannot be identified with an observable taxa.  To deal with this absence, we assume that the reads arise from unobserved haplotypes - the \emph{lineages} - with variation appearing either from mutations along a coalescent genealogy or from errors in SNP ascertainment, informatics, or sequencing.  We take each pool be a mixture of lineages and, conditional upon their number, employ a Bayesian approach to jointly estimate the lineages, mixture proportions, and phylogeny from the SNP read count data.  Since the number of lineages is not known \emph{a priori} we employ an empirical Bayes factor analysis to infer the number of lineages \citep{Newton1994,Kass1995}.  

We assume that short-read sequence data are collected from $N$ pools, indexed by $i = 1, \cdots, N$.  Pools may be the result of differing collection times, spatial locations, or other experimental distinctions.  From the full set of sequence reads, we infer a set of $M$ SNPs, indexed by $j = 1, \cdots, M$.  This may be done by using mapping reads to a reference genome \citep{Li2009} or by employing \emph{de novo} approaches \citep{Zerbino2008, Iqbal2012}.  We suppose that SNPs are biallelic and that counts, $d$, are made for each SNP in each allele state within each pool.  The full data set comprises $\mathcal{D} = [d_{ijs}]$, where $i = 1,\cdots N$, $j = 1, \cdots, M$ and $s \in \{0,1\}$.  Arbitrarily, we assign $s=0$ to be the reference allele state.  Lastly, we assume that the pools constitute independent samples from each other and that changes among SNPs are also independent.  The independence assumptions are computationally expedient but may neglect some useful information, such as linkage or correlations among pool proportions.

\begin{figure}[!ht]
\begin{center}
\includegraphics[scale=0.7]{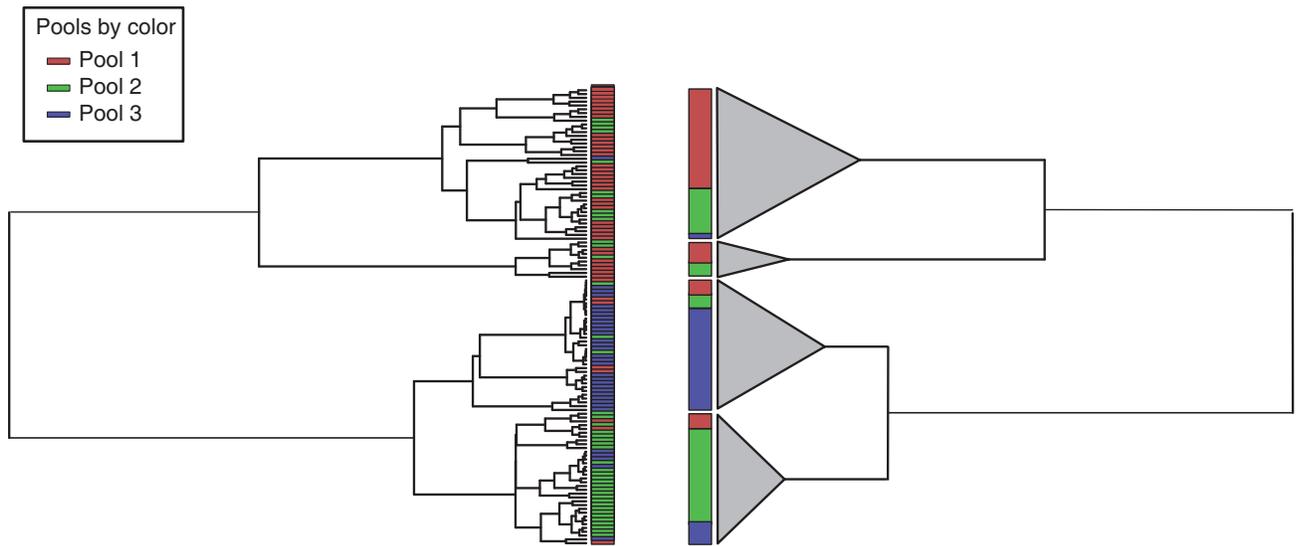}
\end{center}
\caption{A diagram of the lineage model.  On the left hand side, a coalescent process leads to a complete genealogy, with the tips marked by pool as colors.  The right hand side diagrams the lineage model approximation, showing deep branching events together with cones shading the SNP variation indistinguishable from noise.}
\label{fig:lineage_model}
\end{figure}

Our model links two components to provide a likelihood for the SNP count data.  The first piece specifies the structure of SNP variation leading to a set of lineages.  The second piece details the proportions of lineages found in each pool, as in Figure \ref{fig:lineage_model}.  We now lay out each of these components and show how to combine them.  We conclude by detailing the full posterior decomposition from these components and corresponding priors.  Our model has a large number of parameters so we provide a listing of their definitions in Table \ref{table:symbols}. \\

\noindent \textbf{\emph{\small SNP VARIATION}} 

\noindent  We fix number of lineages to be $K$, and number them $k = 1, \cdots, K$.  We assume that there is a rooted coalescent tree, $\mathcal{T}$, specified by a topology, $\tau$, and set of branch lengths, $\{t_b\}$.  By assumption, $\mathcal{T}$ has $K$ external taxa and each corresponds to a lineage, $\mathcal{L}_k$, that defines a haplotype for the SNPs at that tip.  We write out lineages as $\mathcal{L}_k = [l_{kj}]$ where $j = 1, \cdots, M$ and $l_{kj} \in \{0,1\}$ specifies the state of SNP $j$.  The collection of lineages we write as $\mathcal{L}$.   

Since we take SNPs to be independent of each other, we can specify the model for a single SNP without a loss of generality. We suppose that variation in SNP state arises in one of two ways: through mutation along the genealogy, or through some form of observational error.  While errors may arise from a variety of sources including sequencing errors, a poor-quality reference genome, alignment errors or other informatic issues,  we treat them as a resulting from a single homogeneous process. The model consequently loses some power by treating genuine variation that has not reached sufficient presence in the population as an error.     

The model categorizes SNP positions into these two classes, with SNPs arising from mutations on the reduced phylogeny called phylogenetics SNPs, and other SNPs, associated with observational errors, called null SNPs. Of course, phylogenetic SNPs can also be subject to errors but they are not the sole cause of their appearance in the data.  We assume that the type of SNP variation at a site occurs as a Bernoulli trial with a parameter $\lambda$ setting the probability of being a phylogenetic SNP.  This naturally partitions the count data, $\mathcal{D}$, into a phylogenetic component, $\bar{\mathcal{D}}$, and a null component, $\tilde{\mathcal{D}}$.  We refer to this partition by $\mathcal{P}$.  

For each phylogenetic SNP $j$ the allele state for each of the $K$ tips is given by $\mathcal{L}_j = [l_{j 1},\cdots,l_{j K}]$. In a typical phylogenetic context, $\mathcal{L}_j$ would correspond to the observed sequence pattern at a single site in an alignment. Given a mutation rate, $\xi$, we calculate $\mathbb{P}(\mathcal{L}_j|\mathcal{T},\xi)$ using a two-state analog of Jukes and Cantor's mutational model together with Felsenstein's tree pruning algorithm \citep{Jukes1969,Felsenstein1981}.  Each null SNP exhibits an absent pattern across the lineages, with either $\mathcal{L}_j = [0,\cdots,0]$ or $[1,\cdots,1]$.  We assume the probability of either null pattern is $\frac{1}{2}$. \\

\noindent \textbf{\emph{\small POOL PROPORTIONS}}  

\noindent We label the specification of proportions for each lineage in each pool by $\mathcal{S}$.  As each pool is an an exclusive mixture of different lineages, it is natural to capture this structure by an $N \times K$ matrix with each entry $s_{ik}$ giving the proportion of lineage $k$ that is found in pool $i$, enforcing that $\sum_{k=1}^K s_{ik} = 1$ for all $i = 1, \cdots, N$.  \\

\noindent \textbf{\emph{\small LIKELIHOOD}}  

\noindent Supposing that the data are error free, we can relate $\mathcal{L}$ and $\mathcal{S}$ to the data $\mathcal{D}$ in the following way.  Summing over the lineages at each position combines the pool proportions and SNP state to give the expected reference allele frequency for pool $i$ and SNP $j$:
\begin{eqnarray}
p_{ij} & = &\sum_{k=1}^K s_{ik} \cdot (1-l_{kj}) \mbox{.}
\label{snp_prob}
\end{eqnarray}
We assume that sequencing errors afflict all read counts homogeneously with probability $\eta$.  Consequently, we expect only $(1-\eta)$ of the reference counts to come from reference states while $\eta$ of the non-reference counts reflect genuine reference states.  To account for these errors, we correct the reference allele frequency in Equation \ref{snp_prob} by 
\begin{eqnarray}
\tilde{p}_{ij}  &=& (1-\eta)\cdot p_{ij} + \eta \cdot (1-p_{ij}) \nonumber \\
&=& p_{ij} - 2 \cdot \eta \cdot p_{ij} + \eta \mbox{.}
\label{snp_err}
\end{eqnarray}
As SNPs and pools are assumed to be independent, the counts within each pool for each SNP follow a binomial distribution with proportion $\tilde{p}_{ij}$.  This gives the likelihood for the data $\mathcal{D}$ as 
\begin{eqnarray}
\mathbb{P}(\mathcal{D}| \mathcal{L}, \mathcal{S},\eta) & = & \displaystyle\prod_{i=1}^N \prod_{j=1}^M {d_{ij0} + d_{ij1} \choose d_{ij0} } \cdot \Bigg(\tilde{p}_{ij}\Bigg)^{d_{ij0}}\cdot \Bigg(1-\tilde{p}_{ij}\Bigg)^{d_{ij1}}\mbox{.} 
\label{likelihood}
\end{eqnarray} 
	 \\


\noindent \textbf{\emph{\small BAYESIAN INFERENCE}}  

\noindent We can now examine the full posterior decomposition in order to complete our model specification.  Bayes' theorem provides 
\begin{eqnarray*}
\mathbb{P}(\mathcal{L}, \mathcal{S}, \mathcal{P}, \mathcal{T},  \xi,  \eta, \lambda | \mathcal{D}) &\propto& \mathbb{P}(\mathcal{D} | \mathcal{L}, \mathcal{S}, \mathcal{P}, \mathcal{T}, \xi, \eta,\lambda) \cdot \mathbb{P}(\mathcal{L}, \mathcal{S}, \mathcal{P}, \mathcal{T}, \xi, \eta, \lambda) \\
 &\propto& \mathbb{P}(\mathcal{D} | \mathcal{L}, \mathcal{S},\eta) \cdot \mathbb{P}(\mathcal{L}, \mathcal{S}, \mathcal{P}, \mathcal{T}, \xi, \eta, \lambda) \mbox{.} 
\end{eqnarray*}

Noting that $\mathcal{S}$ is independent of all of the other variables and that, conditional upon the partition, $\lambda$ does not affect the lineages, we may then collapse the right-hand side above to be
\begin{eqnarray}
\mathbb{P}(\mathcal{L}, \mathcal{S}, \mathcal{P}, \mathcal{T}, \xi, \eta, \lambda) & = &\mathbb{P}(\mathcal{L} | \mathcal{P},\mathcal{T}, \xi) \cdot \mathbb{P}(\mathcal{P}, \mathcal{T}, \xi, \eta, \lambda)  \cdot \mathbb{P}(\mathcal{S})\mbox{.} 
\label{decomposition_1}
\end{eqnarray}

We first consider the conditional probability for $\mathcal{L}$ in Equation \ref{decomposition_1}.  Since SNPs are independent, we can decompose via $\mathcal{P}$ whether a SNP follows the phylogenetic model or the null model:  
\begin{eqnarray}
\mathbb{P}(\mathcal{L} | \mathcal{P}, \mathcal{T}, \xi) &=& \bigg( \prod_{j \in \bar{\mathcal{D}}} \mathbb{P}(\mathcal{L}_j | \mathcal{T}, \xi) \bigg) \cdot \bigg(\frac{1}{2}\bigg)^{|\tilde{\mathcal{D}}|}  \mbox{,} 
\label{decomposition_2}
\end{eqnarray} 
where $|\tilde{\mathcal{D}}|$ denotes the number of SNPs contained in $\tilde{\mathcal{D}}$.  

We now examine the joint probability in the middle of the right hand side of Equation \ref{decomposition_1}.  Except the partition $\mathcal{P}$ and the parameter $\lambda$, we note that all of the components are independent leading to the relatively simple expression
\begin{eqnarray*}
 \mathbb{P}(\mathcal{P},  \mathcal{T}, \xi, \eta, \lambda) &=& \mathbb{P}(\mathcal{P}|\lambda)  \cdot \mathbb{P}(\mathcal{T}) \cdot \mathbb{P}(\xi) \cdot \mathbb{P}(\eta) \cdot \mathbb{P}(\lambda) \mbox{.}
\end{eqnarray*}
Since a series of Bernoulli trials with parameter $\lambda$ creates the partition, its probability is given by
\begin{eqnarray*}
\mathbb{P}(\mathcal{P} | \lambda)  &=&  \bigg(\lambda\bigg)^{| \bar{\mathcal{D}} |} \cdot \bigg(1-\lambda\bigg)^{|\tilde {\mathcal{D}} |} \mbox{.}
\end{eqnarray*}
With these components specified, we only have to detail the prior distributions, $\mathbb{P}(\mathcal{T})$, $\mathbb{P}(\mathcal{C})$, $\mathbb{P}(\mathcal{S})$, $\mathbb{P}(\xi)$, $\mathbb{P}(\eta)$, and $\mathbb{P}(\lambda)$. \\


\noindent \textbf{\emph{\small PRIOR SPECIFICATIONS}}  

\noindent $\bullet \ \mathcal{S}$ -- We assume that each of the pools is sampled independently from the same prior distribution, so the prior distribution over all the pools is a product of the prior on each.  As we have the constraint that $\sum_{k=1}^K s_{ik} =1$, a natural prior for each pool is a uniform Dirichlet distribution of length $K$, following \citep{Balding1995}.  The prior distribution for $\mathcal{S}$ is then
\begin{eqnarray*}
\mathbb{P}(\mathcal{S}) & = & \prod_{i=1}^N \mbox{\small{DIRICHLET}}(s_{i1},\cdots, s_{iK} | \mathbf{1}_K)\mbox{,}
\end{eqnarray*}
where $\mathbf{1}_K$ is a vector of ones of length $K$ \citep{Pritchard2000}.

\noindent $\bullet \ \mathcal{T}$ --  We assume a coalescent prior for $\mathcal{T}$.  If $\{u_i : i = 2, \cdots, K \}$ are the time intervals between coalescent events ordered to reflect the number of individuals present at that time then the tree has total branch length $T = \sum_{i=2}^K i \cdot u_i$ and $$\mathbb{P}(\mathcal{T}) = \prod_{i=2}^K e^{-{i \choose 2}\cdot u_i}\mbox{.}$$  The distribution for the total branch length $T$ can be found in \citet{Tavare1983}.\\
\noindent $\bullet \ \xi$ --  This is distributed as $\mbox{Exp}(1)$.  \\
\noindent $\bullet \  \eta, \ \lambda$ -- We assume these are uniform on the open unit interval, $(0,1)$. \\

\subsection{Inference}
We use a Metropolis-Hastings Markov chain (MCMC) approach to inference.  In order to infer the parameters $\mathcal{S}, \mathcal{T}$, $\mathcal{L}$, and $\mathcal{P}$, we employ approaches previously applied to phylogenetics \citep{Huelsenbeck2001}.  To infer $K$ we use an empirical Bayes factor procedure that integrates information across a set of MCMC runs.  Conditional upon a fixed $K$, we now describe the parameter updates.

The Metropolis-Hastings ratio gives the probability that a proposed parameter update $x'$ will be accepted from a current state $x$ with probability $\alpha$ such that $$ \alpha = \mbox{min}\bigg( \frac{\mathbb{P}(x')}{\mathbb{P}(x)}\cdot \frac{ \mathbb{P}(x' \rightarrow x) }{ \mathbb{P}(x \rightarrow x') },1 \bigg) = \mbox{min}\bigg( \alpha_1 \cdot \alpha_2, \ 1 \bigg)\mbox{.}$$
The first fraction is the ratio of the posterior probability of $x$ and $x'$, and we denote this $\alpha_1$.  The second is the ratio of the probability of choosing the current state from the proposed state over the reverse move.  We label this $\alpha_2$.  Since $\alpha_1$ constitutes assessment of the likelihood and the prior functions which can be calculated as shown above, we subsequently only consider $\alpha_2$.  

\subsubsection{$\mathcal{T}$}
For each iteration, we propose a subtree prune and regraft (SPR) move \citep{Felsenstein2004}.  As the tree is rooted, a node is chosen uniformly among all nodes within the tree not connecting above to the root.  Removing this node divides the topology $\tau$ into $\tau_p$, the pruned segment, and $\tau_r$, the remaining segment.  We re-attach $\tau_p$ to $\tau_r$ along an uniformly chosen edge within $\tau_r$, with the precise location taken uniformly across the chosen edge.  This generates a new tree $\tau'$ and corresponding branch lengths $\{t_b^*\}$.  We then recalculate the branch lengths to ensure a coalescent tree.  Since the starting and ending states are equally probable with respect to each other, $\alpha_2=1$.   To ensure the chain does not get stuck in a mode of the posterior distribution, we also propose new branch lengths by successively proposing small changes in length to each $t_b$ on a uniform interval $[t_b - \epsilon, t_b + \epsilon]$.  For both moves the probability of proposal is the same in both directions so $\alpha_2 =1$.  

%

\subsubsection{$\mathcal{L}$ and $\mathcal{P}$}
The inference of $\mathcal{L}$ for a given SNP $j$ involves a simple bit-flip operation.  First, a SNP $j$ and a lineage $k$ are selected at random and allele state for that lineage's SNP is flipped: $l_{kj} \rightarrow |1 - l_{kj}|$.  Since this a deterministic operation, and the SNP and lineage are chosen uniformly, $\alpha_2 =1$.  It is not necessary to infer directly $\mathcal{P}$, since SNPs with site patterns that are uniform -- where all allele states are $0$ or $1$ -- are treated null positions, while those that are not uniform are treated as phylogenetic.

\subsubsection{$\mathcal{S}$}
We update $\mathcal{S}$ by the composition of a randomly chosen pool $i$.  We propose a new pool $S_i'$ by drawing from a Dirichlet distribution with parameters $(\gamma_1, \cdots, \gamma_K)$ informed by $\mathcal{R}$ such that
\begin{eqnarray*}
\gamma_k &= &1 + \beta \cdot \frac{\sum_{j = 1}^M (1 - l_{jk}) \cdot d_{ij0} + l_{jk} \cdot d_{ij1}}{\sum_{k= 1}^K \bigg[\sum_{j = 1}^M (1 - l_{jk}) \cdot d_{ij0} + l_{jk} \cdot d_{ij1}\bigg]} \mbox{,}
\end{eqnarray*}
where $\beta$ is a tuning parameter.  In practice, we find $\beta = 5$ to provide good rates of move acceptance.  A brief calculation shows that $\alpha_2 = \prod_{i=1}^K \bigg(\frac{s_{ik}}{s_{ik}'}\bigg)^{\gamma_k-1}$.

\subsubsection{$\eta$, $\lambda$ and $\xi$}
All these are drawn directly from the prior and so have trivial Hastings ratios.  

\subsubsection{$K$}

We run the MCMC for $K = 2, \cdots, 2 \cdot N$ and then compare the runs using Bayes' factors to find $K$.  To infer the Bayes' factor between each pair of runs, we require the marginal likelihood $\mathbb{P}(\mathcal{D} | K)$, and estimate it by taking the harmonic mean of an importance sample from the likelihood using the posterior density as weights, as in \citet{Kass1995}.  This estimator of the marginal likelihood is known to have poor performance in certain circumstances, although we empirically observe it to work well in the simulations below, as it has as in other phylogenetic contexts \citep{Drummond2007,Ronquist2003}.
\section{Simulations}

\subsection{Simulations under the model}

To examine the performance of the model, we simulate data under the model with a variety of parameters and then compare against inferred values.  We simulate coalescent trees with a fixed number of segregating sites using the \verb+ms+ program \citep{Hudson2002} and sample with replacement from the created sequences to get the desired number of SNPs.  We then randomly choose a fraction of these SNPs to be null and set all their allele states to zero or one with probability $\frac{1}{2}$.  We then generate the mixture coefficients by drawing $N$ times from a Dirichlet distribution with $\mathbf{\alpha}_K$ varying with a mixture parameter $\rho$.  We construct $\mathbf{\alpha}_K$ as $\mathbf{1}_K + \rho \cdot \mathbf{1}_{u> 0.1}$, where $\mathbf{1}$ an indicator function and the vector $u$ consists of $K$ uniform draws from the unit interval.  Combining the lineages and pool proportions with a specified error rate as in Equation $\ref{snp_err}$, we draw the sought number of read counts for each SNP from a binomial distribution with parameter $\tilde{p}_{ij}$.  To understand the performance of the algorithm across different parameter regimes we simulated SNP count data with parameters found in Table \ref{table:sim_paras}.  For all parameter values, we fixed the number of pools to $N=7$ and the number of lineages to $K=6$ and ran ten independent iterations.

\subsubsection{An example}
We begin with an in-depth example from the simulations, with $250$ SNPs, a read depth of $10$, $\lambda = 0.95$,  $\eta = 0.001$, and $\rho = 4$. We select an iteration where the model moderately underestimates the number of lineages in order to examine how the model copes with partially incorrect inference.  

\begin{figure}[ht]
\centering
 	 \subfigure{
	   \includegraphics[scale =0.3]{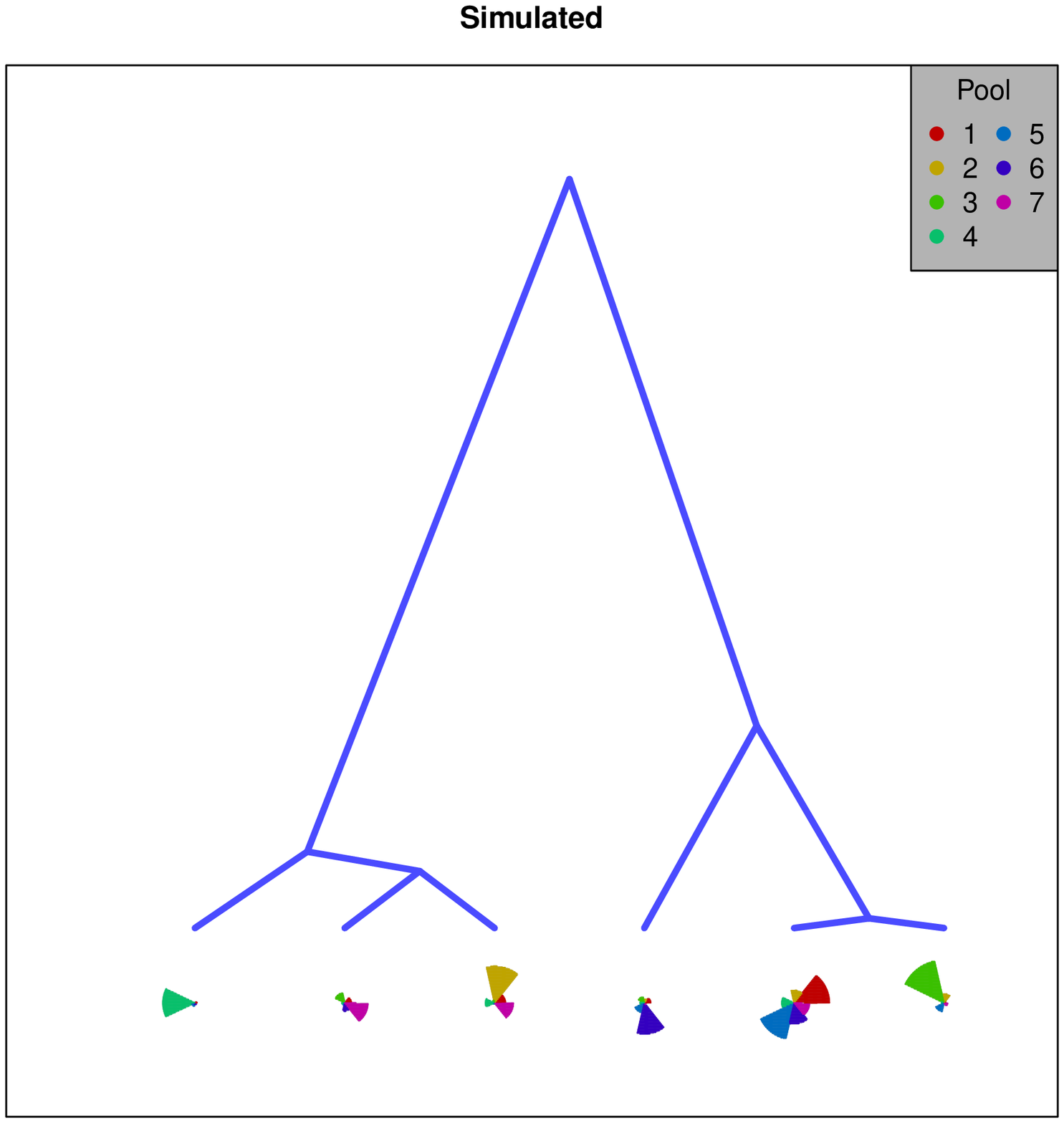}
	 }
	   \subfigure{
	   \includegraphics[scale =0.3]{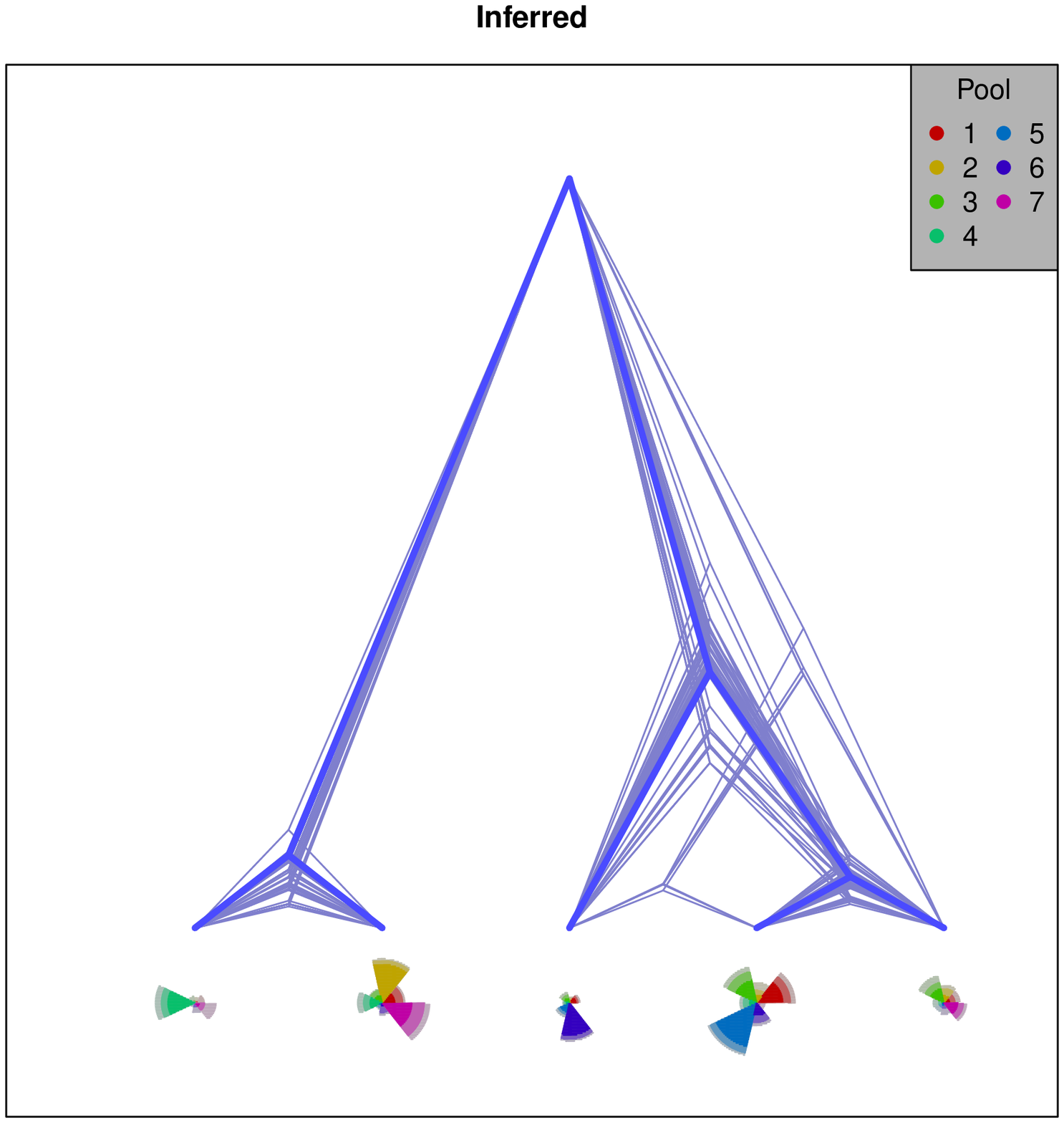}

	 }

	\caption{Comparison between simulated tree and pool proportions (left) and inferred trees and pool proportions (right).}
	\label{sim_inf_ex}
\end{figure}

\begin{figure}[ht]
\centering
	\subfigure{
	   \includegraphics[scale =0.32]{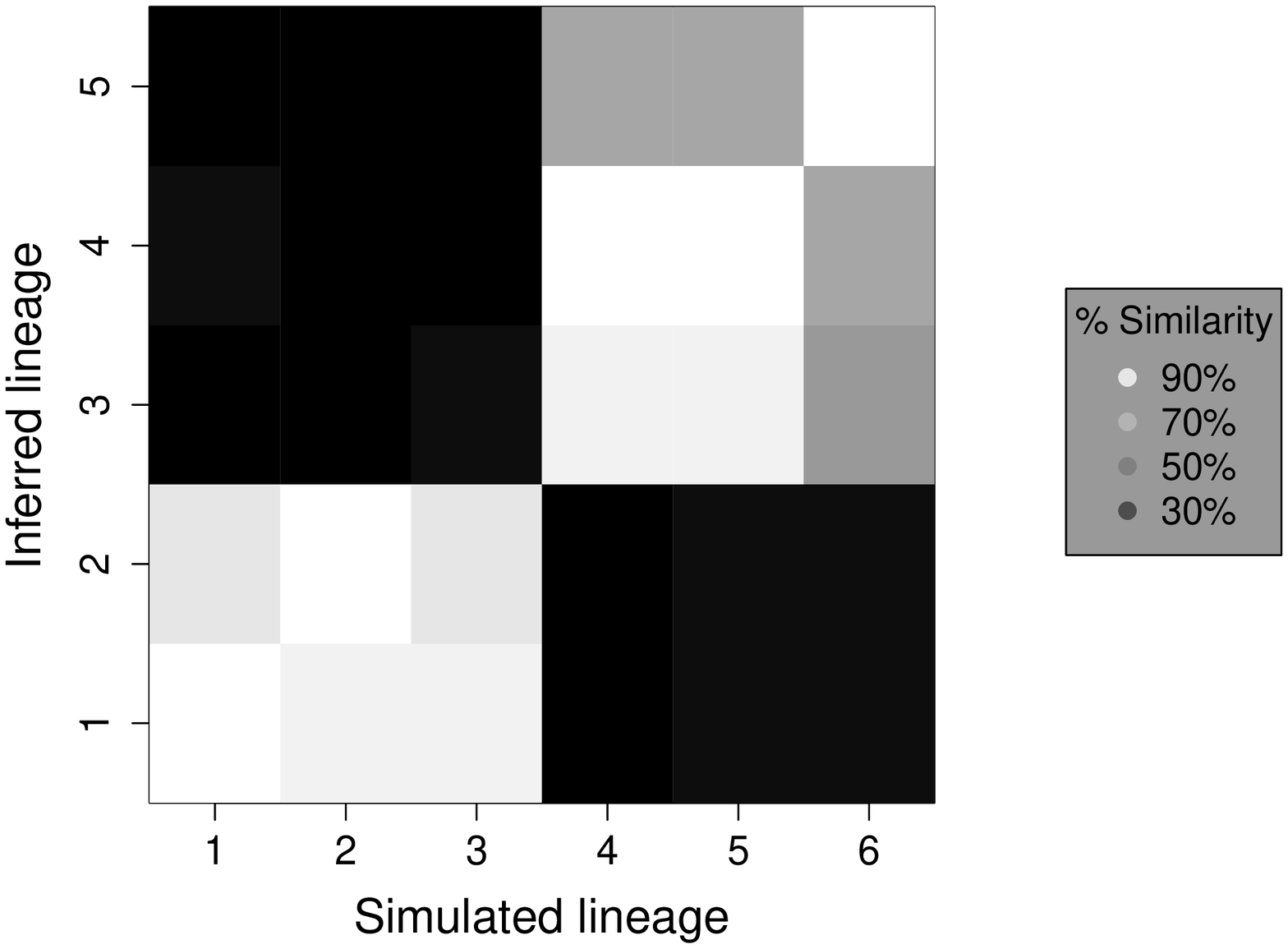}
	
	}
	\subfigure{
	   \includegraphics[scale =0.32]{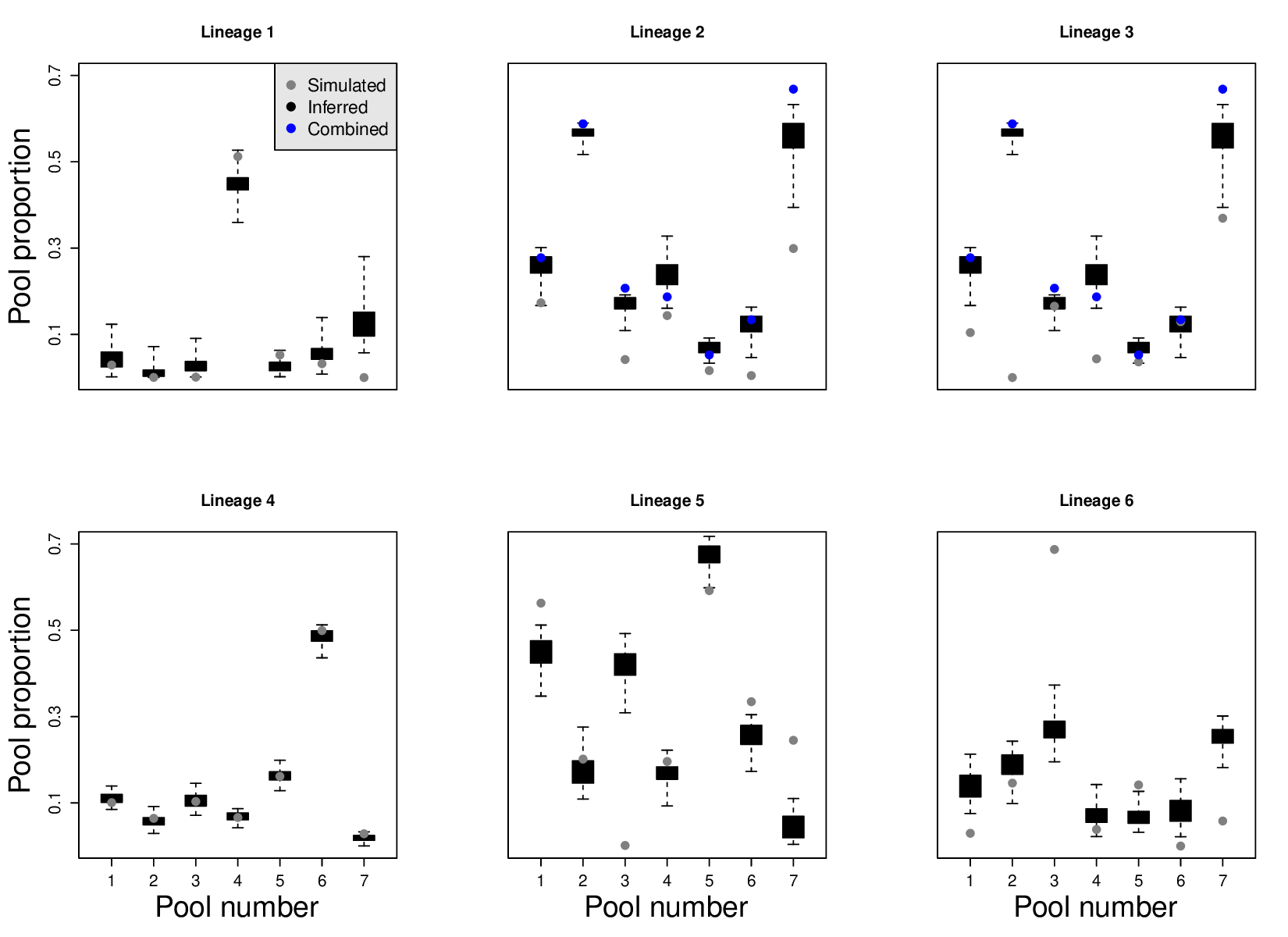}
	}

	\caption{(Left) Percentage of concordant between simulated and inferred lineages. (Right) Comparison between pool proportions for simulated (light grey) and inferred (dark grey) values for each simulated lineage.  Blue dots show combined proportions for simulated lineages $2$ and $3$. }
	\label{line_pool_ex}
\end{figure}

We present the simulated and inferred lineage models in Figure \ref{sim_inf_ex}.  The dark tree shows the maximum posterior probability tree while the remaining trees in light blue each show a sample from the MCMC.  The model infers only $5$ lineages, collapsing lineages $2$ and $3$ into one, although the trees appear otherwise nearly always congruent.  The pie charts of pool proportions below the inferred trees show the $5\%$, mean, and $95\%$ estimates.  The mean estimates appear close to the simulated values, although some fraction of the proportion for lineage $6$ in pool $3$ appears to have `migrated' to lineage $5$.  The left side of Figure \ref{line_pool_ex} compares the SNP patterns of the six simulated lineages against the five inferred lineages, with the lowest fraction of concordance within any column as $83\%$.  The right side of the figure shows that inference of pool proportions performs generally well.  Direct comparison of simulated pool proportions for lineages $2$ and $3$ appears to indicate poor performance, although we observe that combining the simulated values for these lineages (in blue) substantially improves the agreement.  \\

\noindent \textbf{\emph{\small COMPARISON TO PCA}}  

\noindent Absent an explicit modeling framework, researchers might naturally seek to understand metagenomic SNP count data by using principal components analysis (PCA), a general approach to high-dimensional data exploration \citep{Jolliffe2005}.  We compare the results above to those from PCA, as shown in Supplementary Figure 1.  The PCA analysis indicates that a large majority of the variation between samples can be explained by the first two components.  Examination of these components shows a distinct separation of pools $1, \ 4, \ 5, \mbox{ and } 6$ from pools $2,\ 3, \mbox{ and } 7$, consistent with simulated data.  Additional components give similar portraits but with additional separation for pool $6$ from pools $1, \ 4, \mbox{ and } 5$.  In this example PCA analysis appears to provide a general method of separating pools based on SNP count similarity but is difficult to further interpret.

\subsubsection{Comparison across parameters}

\begin{figure}[ht]
\centering
 	 \subfigure{
	   \includegraphics[scale =0.36]{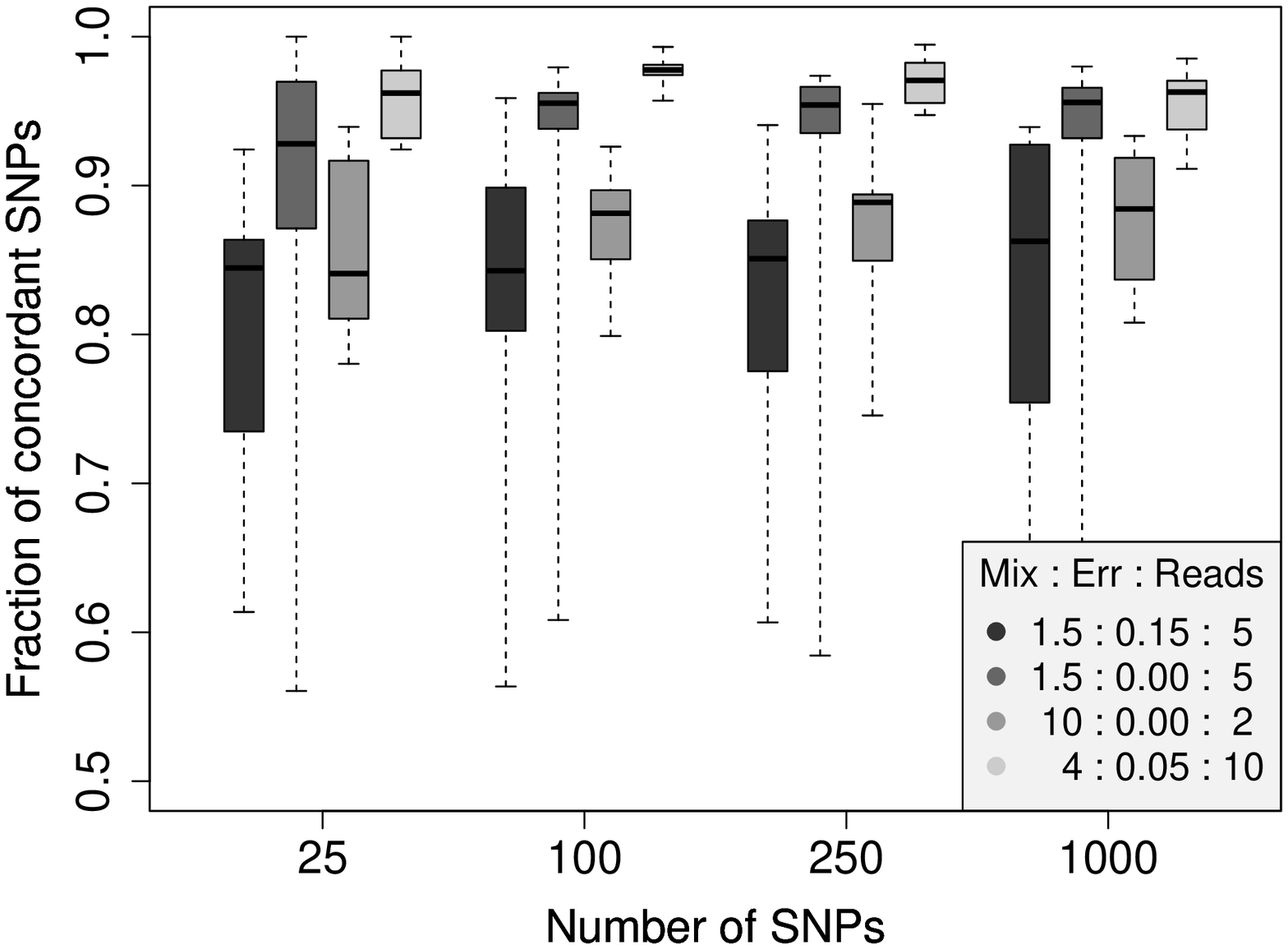}
	 }
 	 \subfigure{
	   \includegraphics[scale =0.36]{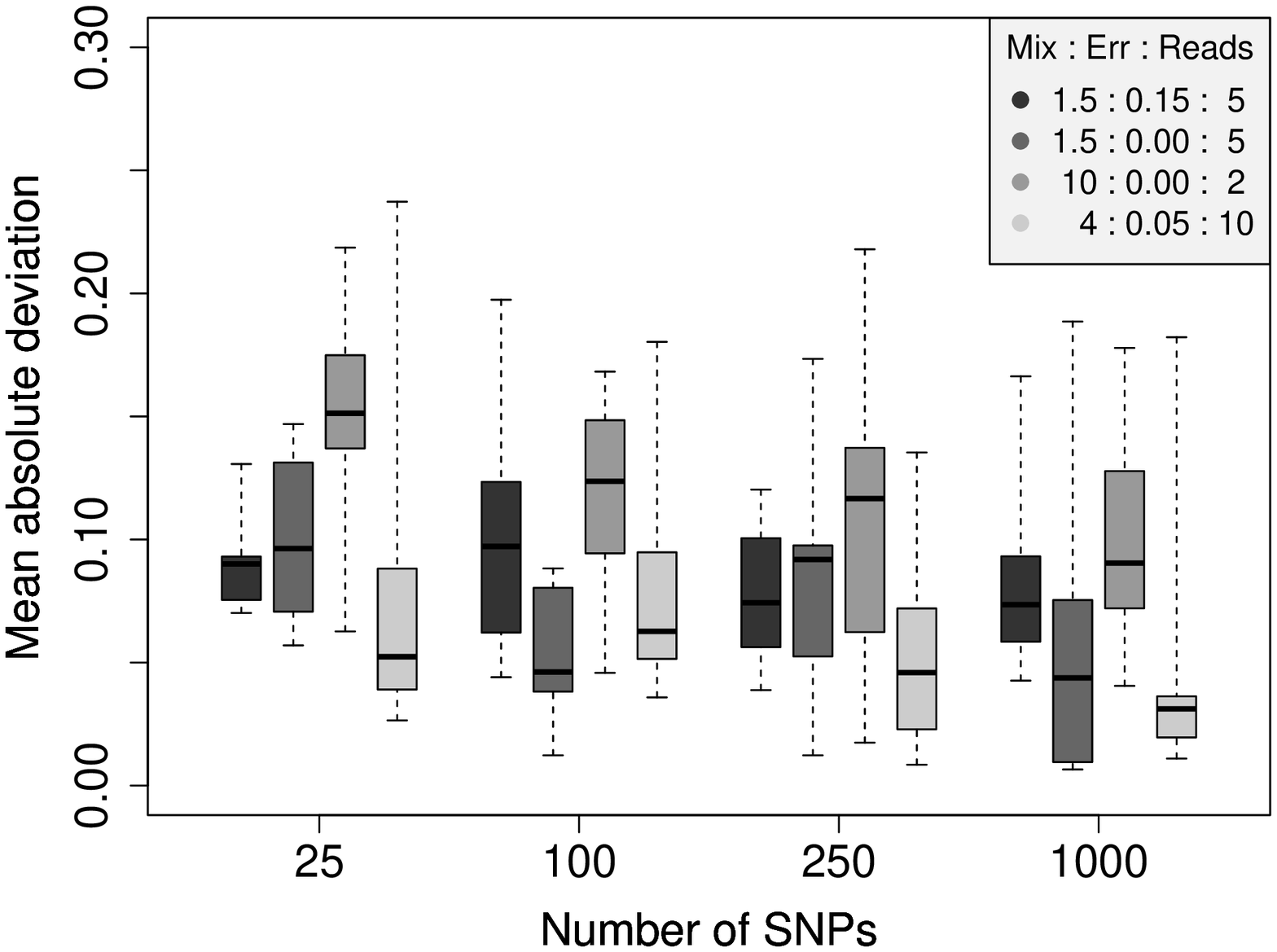}

	 }
	   \subfigure{
	   \includegraphics[scale =0.36]{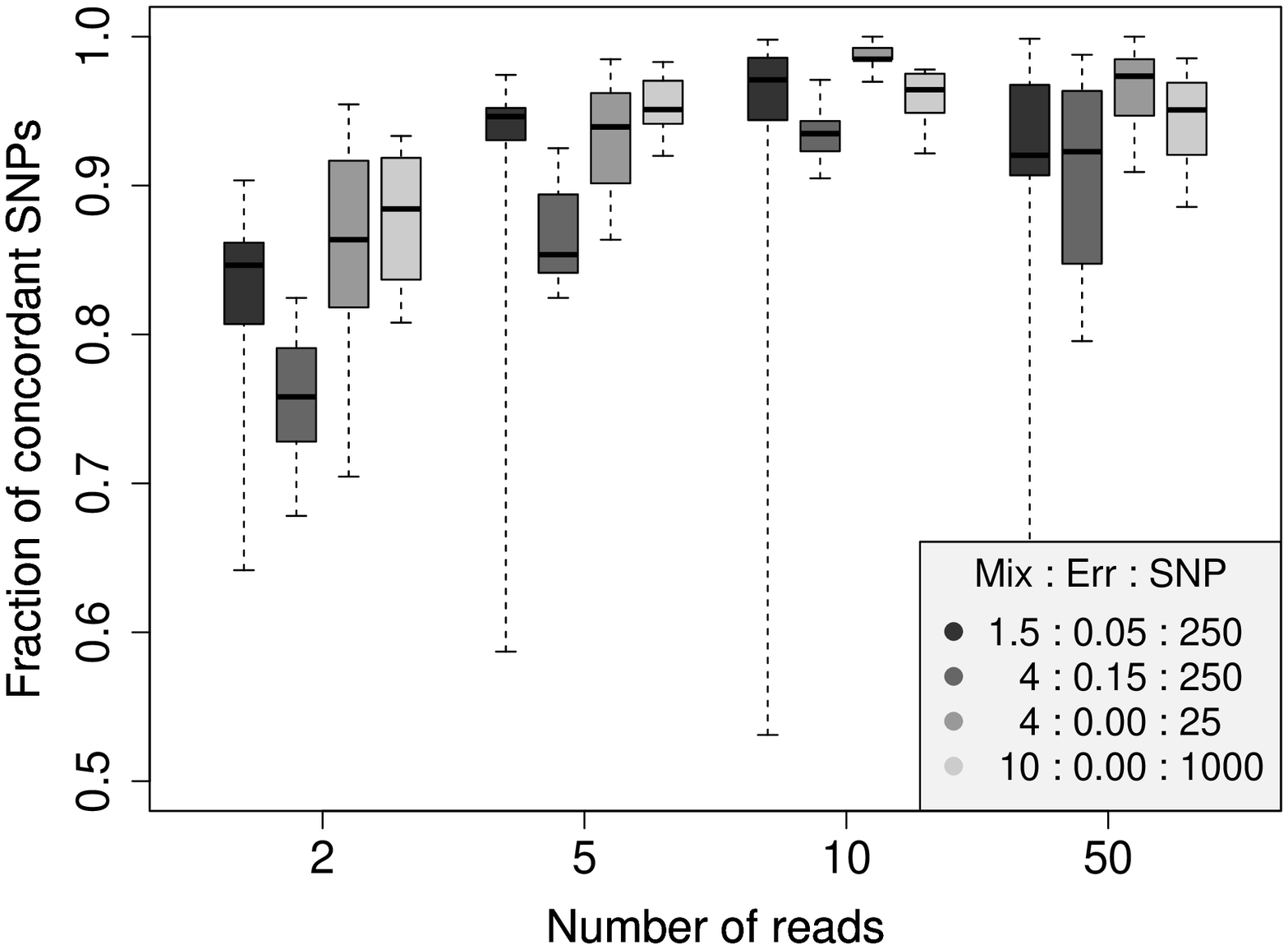}

	 }
	   \subfigure{
	   \includegraphics[scale =0.36]{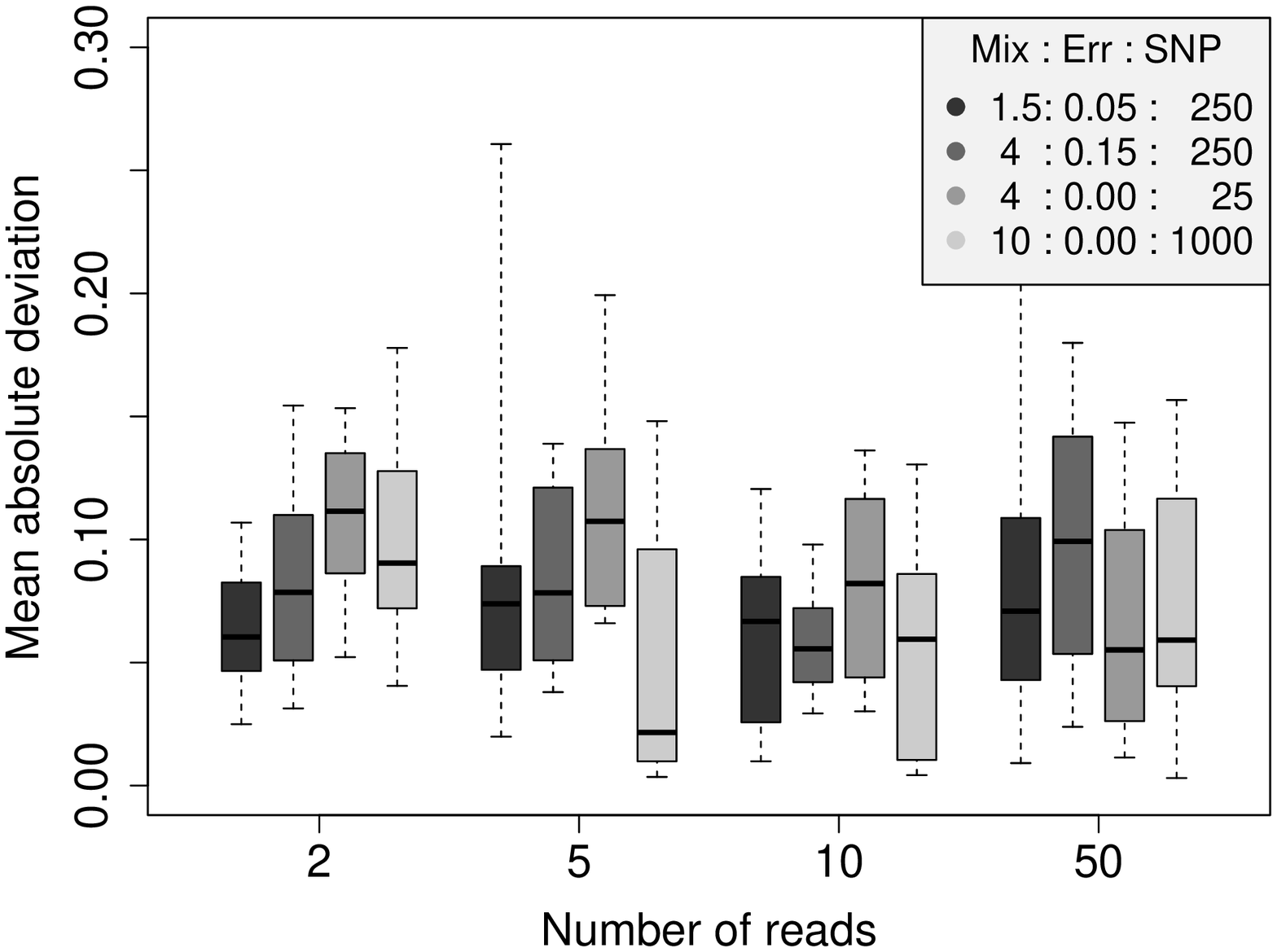}

	 }
 	 \subfigure{
	   \includegraphics[scale =0.36]{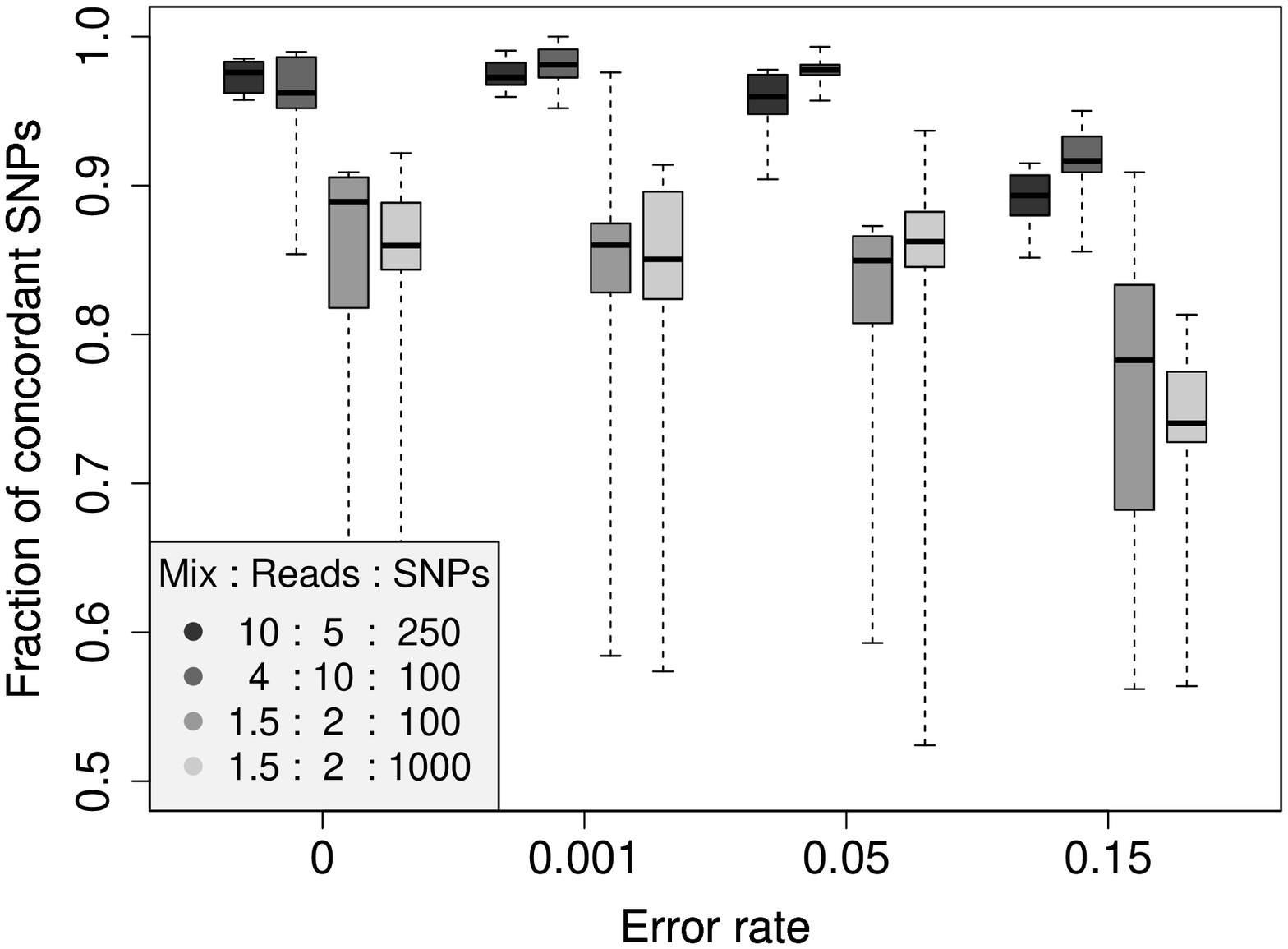}

	 }
 	 \subfigure{
	   \includegraphics[scale =0.36]{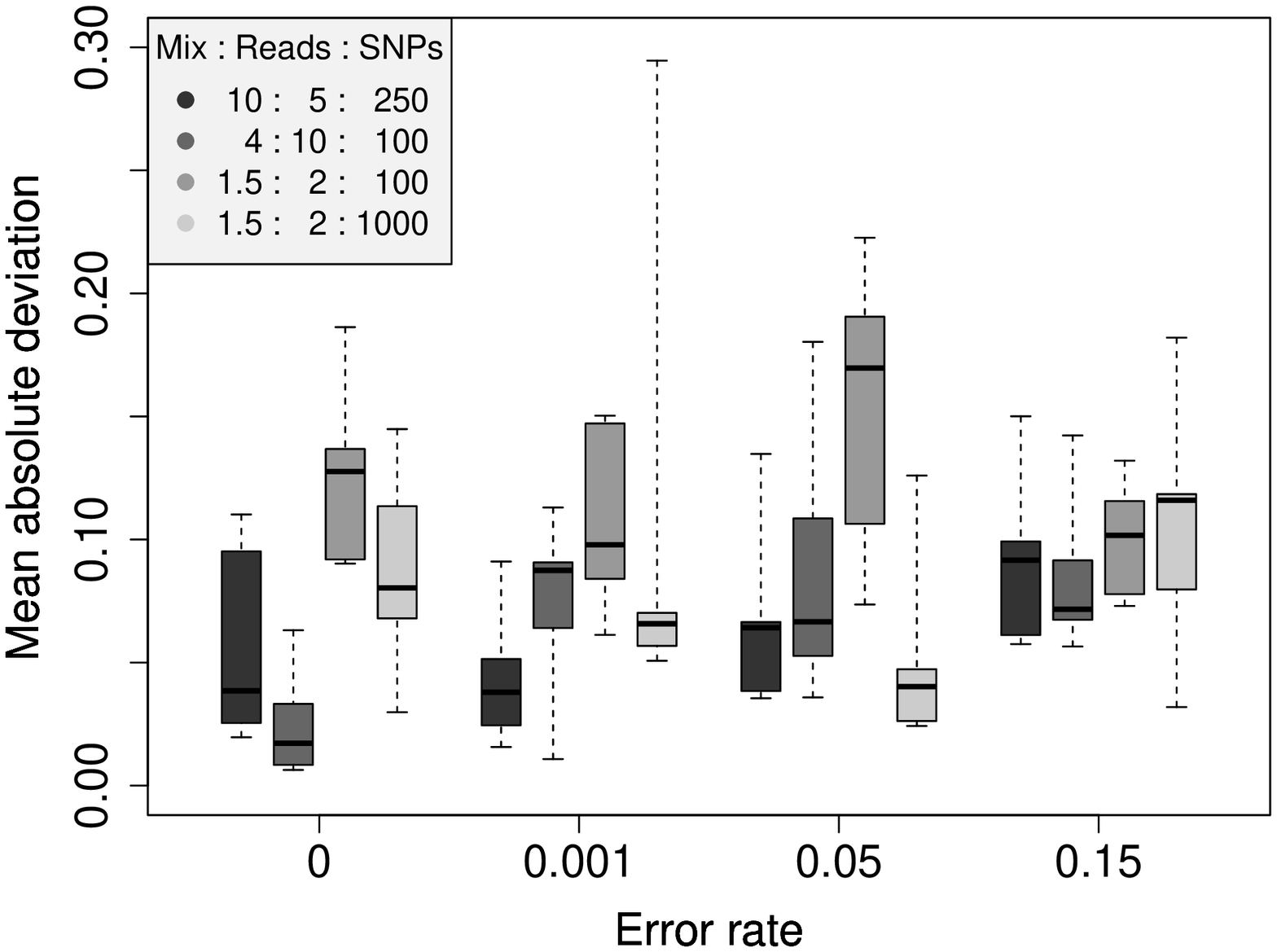}

	 }

	\caption{Comparison of simulated and inferred values for lineages (left column) and pool proportions (right column) by number of SNPs (top row), number of reads (middle row) and error rate (bottom row).  }
	\label{fig:sim_perf}
\end{figure}

We present the collected results for the model simulations in Figure \ref{fig:sim_perf} for varying numbers of SNPs, read count depths, and error rates.  The left column shows lineage performance in terms of the fraction of concordant SNPs between each simulated lineage and its closest inferred lineage.  The right column shows pool performance as the mean absolute deviation between simulated and inferred values.  The summaries indicate that the read count depth affects performance most strongly, with more moderate changes coming from the number of SNPs and the error rate.  The number of SNPs and error rate more strongly influence pool proportion inference, where read count contributes little.  We also find that increasing mixing correlates with increasingly poor lineage concordance (Supplementary Figure 2).  The fraction of null SNPs alters performances negligibly.     

\subsubsection{Topological performance and model selection}

Assessing the topological performance for the lineage model presents a significant challenge due to two related issues: that the number of taxa is not fixed, and that the taxa themselves are not uniquely identifiable.  In standard phylogenetic contexts, the fixed number of samples and their unique identification are implicitly used in standard algorithms to assess topological congruence \citep{Planet2006}.  We have not able to find an applicable approach in the literature nor have we been able to develop a straight-forward extension ourselves.  To provide some understanding of the quality of model performance, we visually examine the output of ten iterations from three parameter regimes: low-quality data  ($M=25$, $\eta = 0.15$, $d=2$, $\rho=1.5$); moderate-quality data ($M=100$, $\eta = 0.05$, $d=5$, $\rho = 4$); and high-quality data ($M=250$, $\eta=0.001$, $d=10$, $\rho=10$).  We find that empirical Bayes factor analysis underestimates the number of lineages in the low-quality regime, as might be expected, but infers values near to the simulated number for the moderate- and high-quality sets, as in Supplementary Figure $3$.  In these latter two cases, we visually compare the inferred tree against the simulated tree and find they are often consistent.  Errors encountered most often took the form of merged lineages or `migrating' pool proportions (Supplementary Figure $4$), and nearest-neighbor interchanges between taxa.  

\subsubsection{Algorithmic performance}

We implement the lineage model in C++ using the GNU Scientific Library.  Our implementation shows reasonable computational speed and convergence for an MCMC-based approach, and is appropriate for thousands of SNPs and up to a hundred pools.  For a set of $1000$ SNPs, $7$ pools, and $6$ lineages, a complete analysis ($2\cdot10^{6}$ MCMC iterations) required slightly more than $10$ hours on a multi-core Linux-based laptop with $2.1$ gigahertz processor.  As a point of comparison, this data has substantially more SNPs than in our empirical examples, and on the same order as publicly available microbiome data.  The algorithm performs linearly in the number of SNPs, the number of pools, and worse than linearly in the number of lineages.  Copies of the code and ancillary scripts are available upon request.  

Using the CODA package in $R$, we apply several standard metrics to assess the convergence of the algorithm, including the Gelman-Brooks test, autocorrelation analysis, and Raftery estimation of burn-in length \citep{Plummer2006,Brooks1998,Geweke1991,Cowles1996,Raftery1992}.  All tests indicate that the MCMC converges rapidly and consistently.  Examination of the Gelman-Brooks statistics and autocorrelation analysis reveals that thinning MCMC chain output to one iteration in a thousand sufficient to provide effective sampling.  The Raftery estimation suggests that $1e6$ iterations are sufficient to achieve stationarity for a test data set with $1000$ SNPs.  For most data sets, we see $10-50\%$ acceptance rates for all parameters.  Observationally, we find the model applied repeatedly on a wide variety of data sets achieves nearly identical parameter estimates.   

\subsection{Simulations under the island coalescent}

To understand the model's performance under a more realistic -- but still idealized -- context, we also simulate polymorphism data under the island coalescent model \citep{Wakeley2001,Hudson2002}.  This model structures a coalescent process by allowing individuals to migrate among segregated populations (islands) at asymmetric rates.  We employ the \verb+ms+ package to simulate the phylogenetic tree, specifying five islands and assuming that the population size is constant at $120$ individuals within each island.  Ideally, we would be able to simulate such that each read comes only occasionally comes from the same individual, as could be expected in a microbial experiment.  Unfortunately, we cannot do this computationally and instead use this finite approximation.  To generate the migration rate matrix we first draw from a Dirichlet distribution with parameters drawn from a Beta distribution with $\alpha=1$, $\beta=4$, and then we multiply all off-diagonal entries by a constant $\psi$ that we call the mixing proportion.  We can then scale the degree of migration among islands, with the limit $\psi = 0$ enforcing the island populations to be fully segregated.  Having generated an appropriate tree, we use $R$ scripts to generate polymorphism data in the following way.  Following the infinite sites model, we distribute SNPs along the branches of the tree with probability proportional to branch lengths.  This specifies the full haplotypes for each of the individuals.  We then sample randomly across all sites and individuals, adjusting the number of each to account for numbers of reads and SNPs \citep{Kimura1969}.  We aggregate the results within islands to generate pool-specific count data.  We use $10$ for read depth and $1000$ SNPs.  

\subsubsection{An example}

\begin{figure}[ht]
\centering
 	 \subfigure{
	   \includegraphics[scale =0.29]{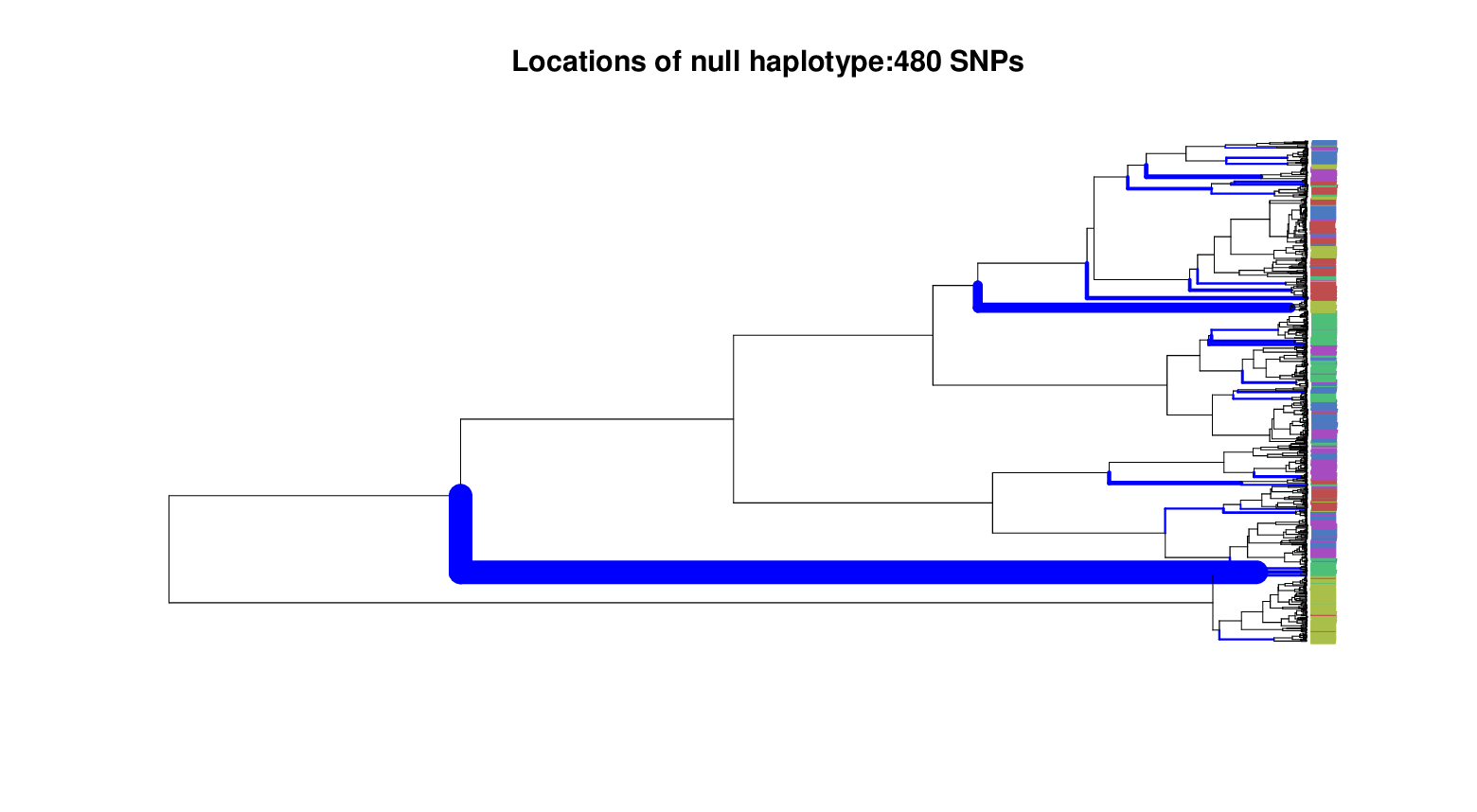}

	 }
	   \subfigure{
	   \includegraphics[scale =0.29]{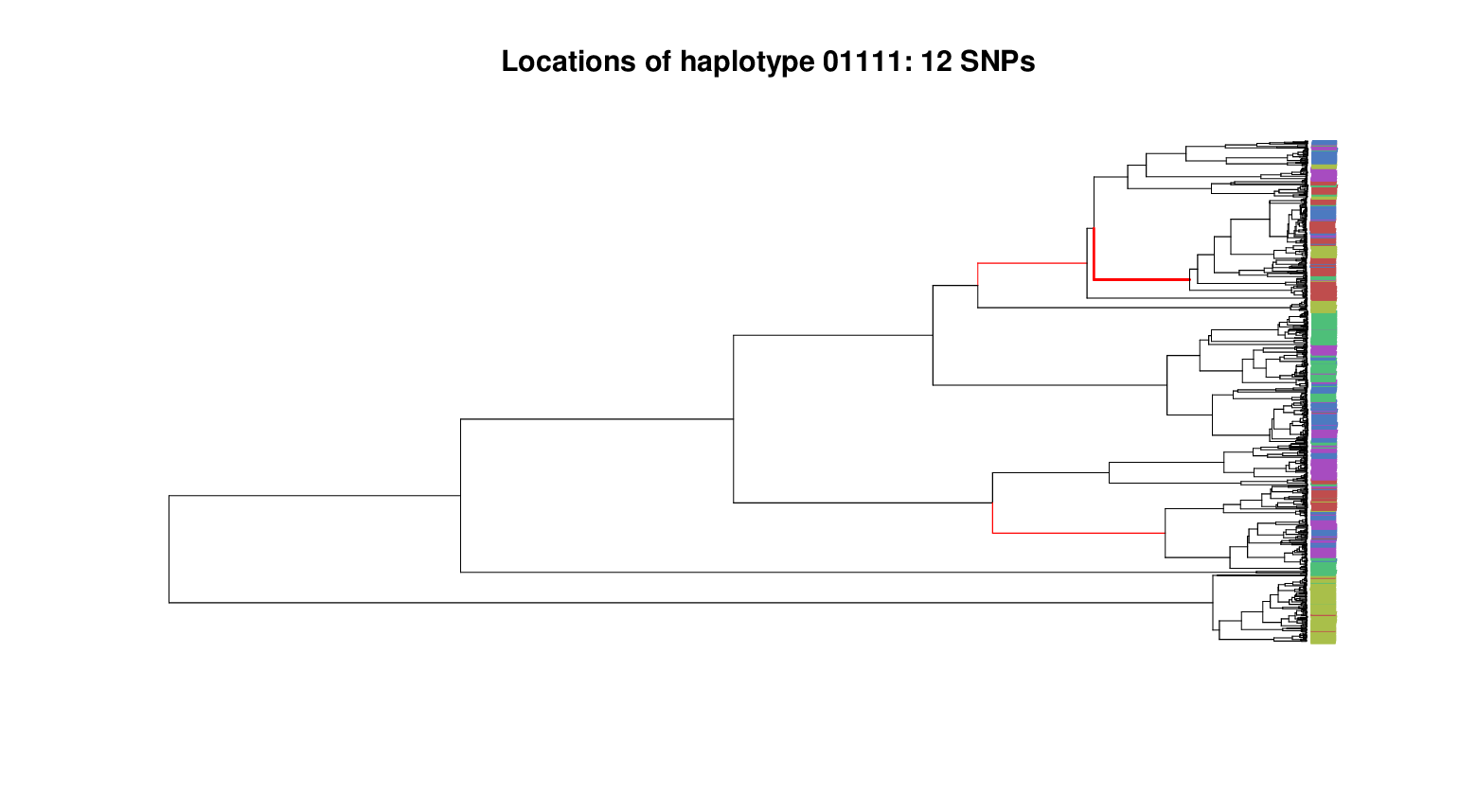}

	 }
 	 \subfigure{
	   \includegraphics[scale =0.29]{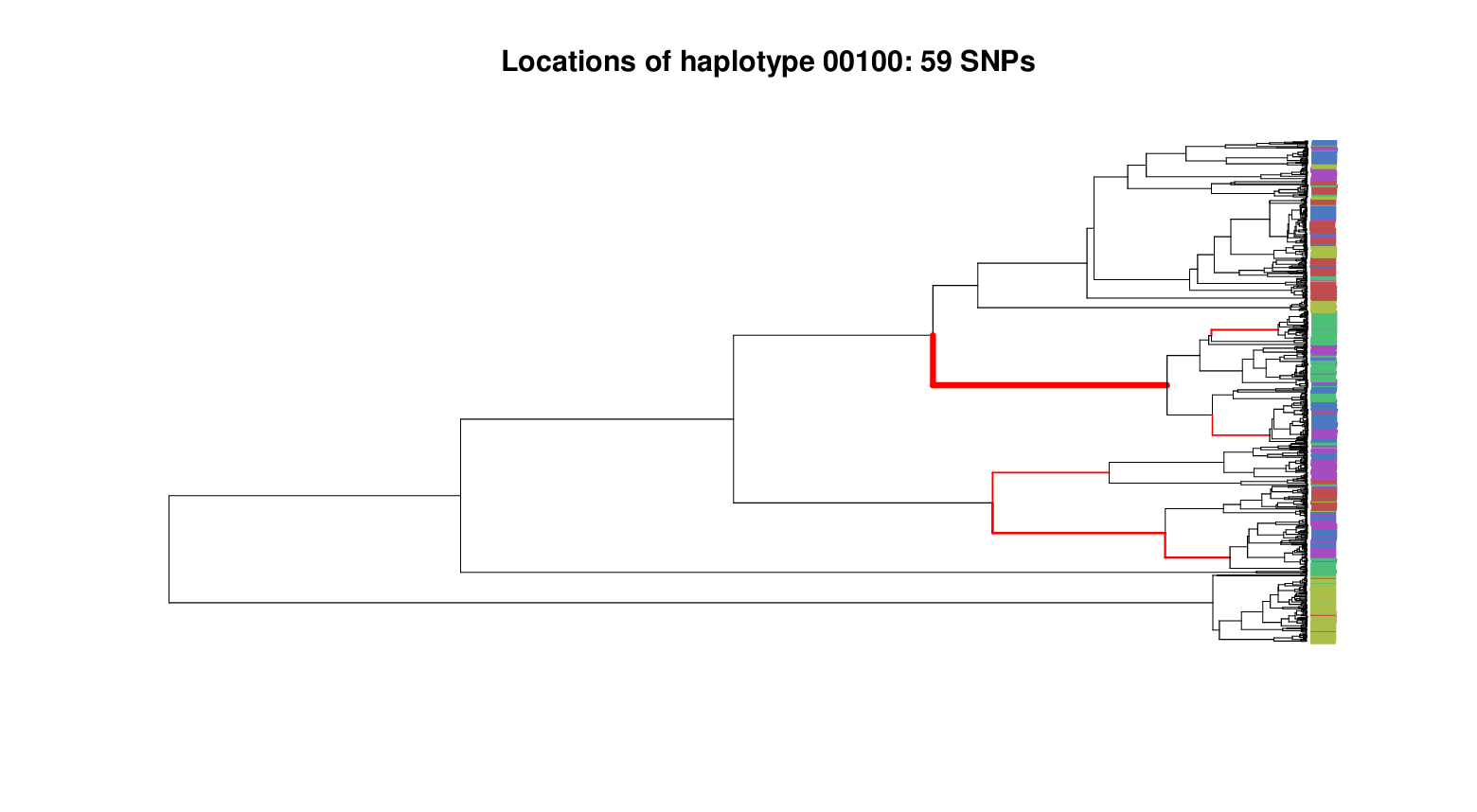}

	 }
 	 \subfigure{
	   \includegraphics[scale =0.29]{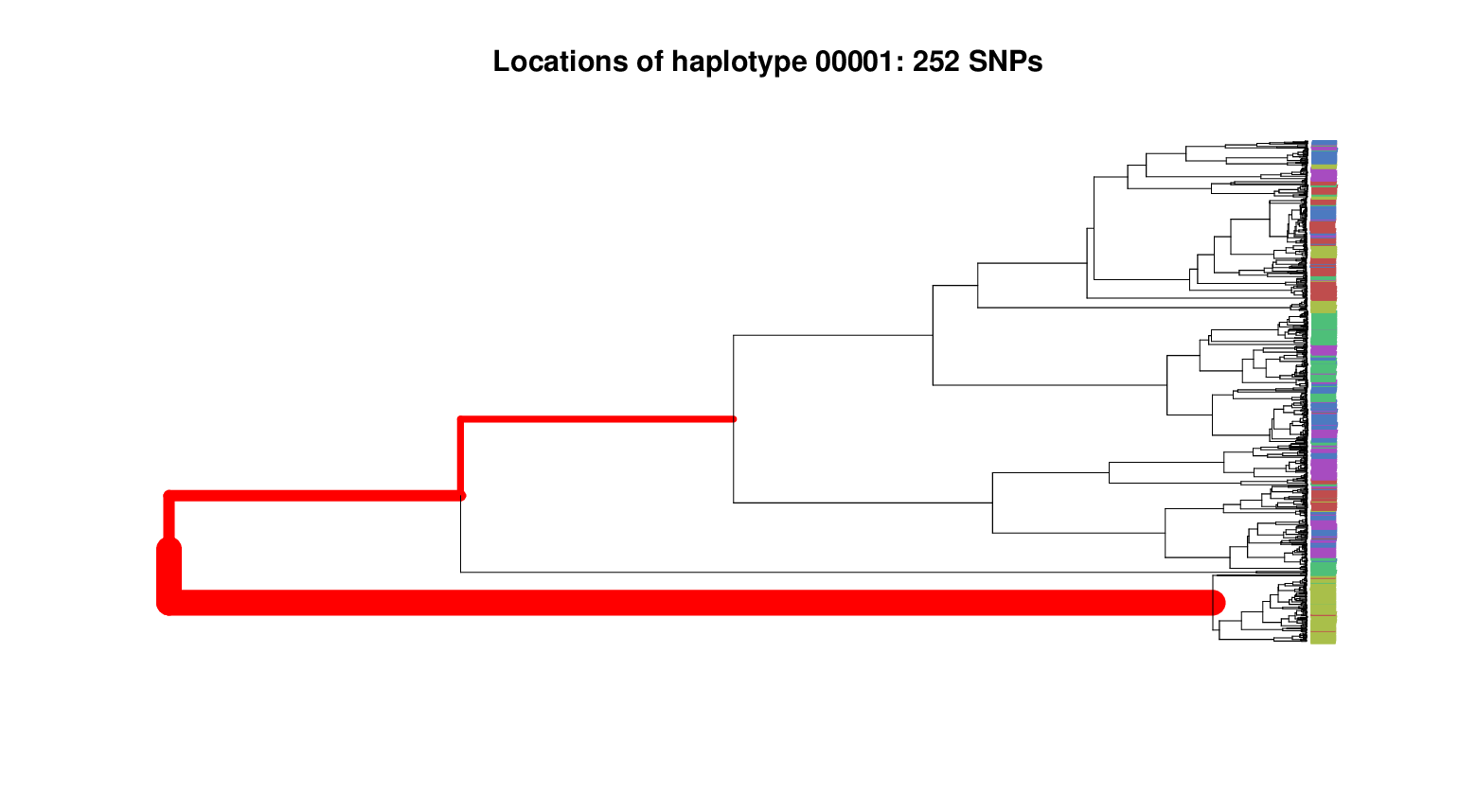}

	 }
 	 \subfigure{
	   \includegraphics[scale =0.29]{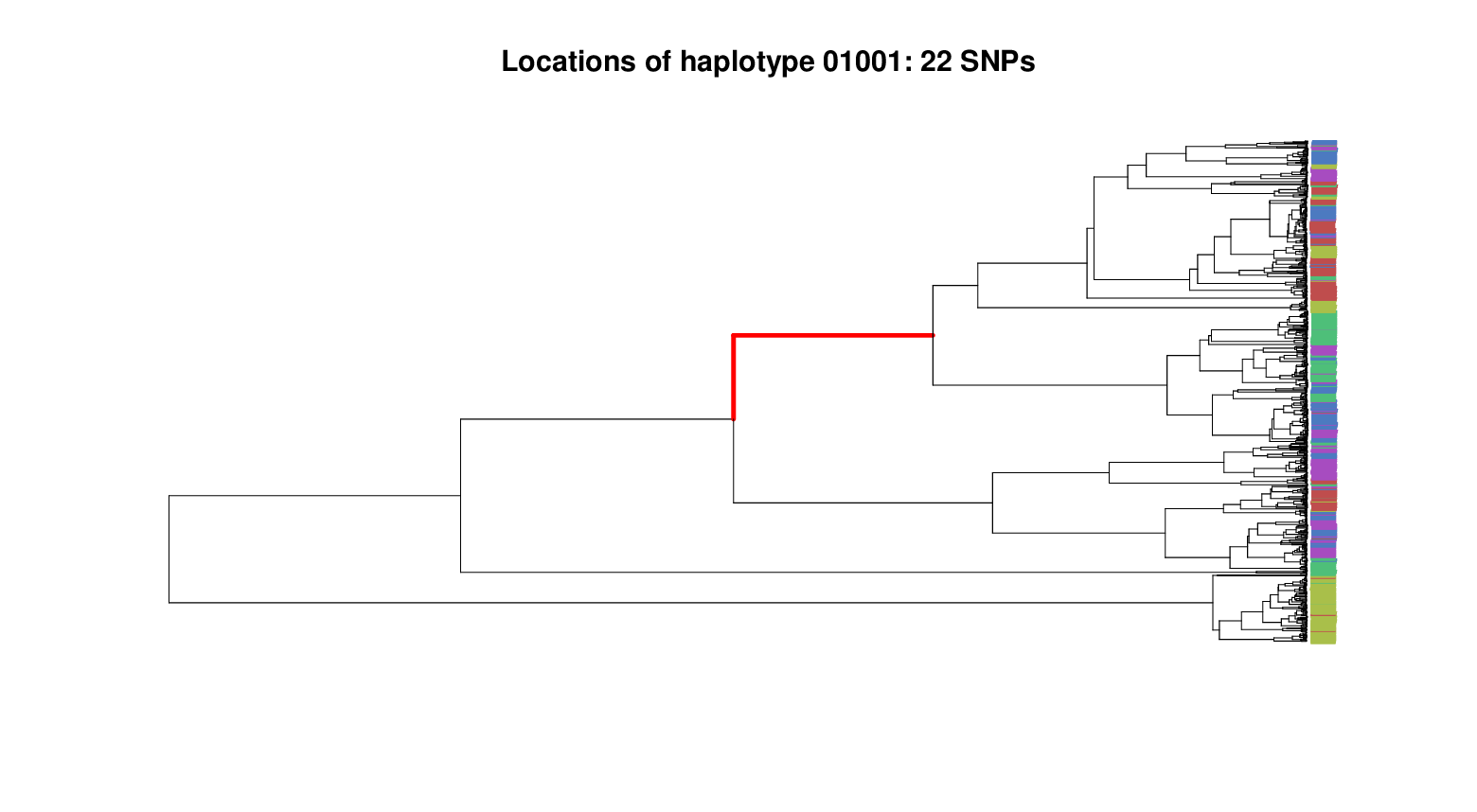}

	 }
 	 \subfigure{
	   \includegraphics[scale =0.29]{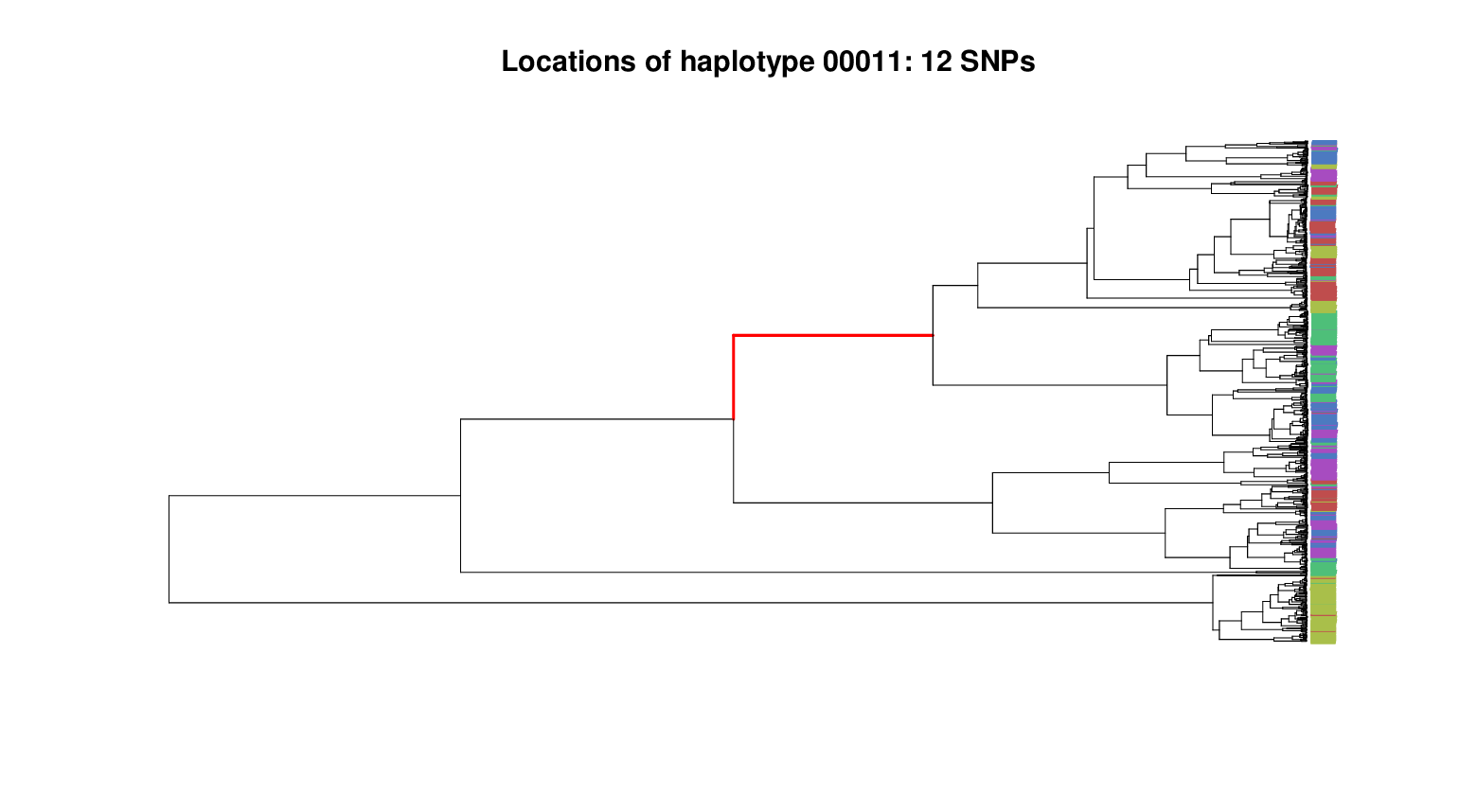}

	 }
 	 \subfigure{
		\includegraphics[scale=0.35]{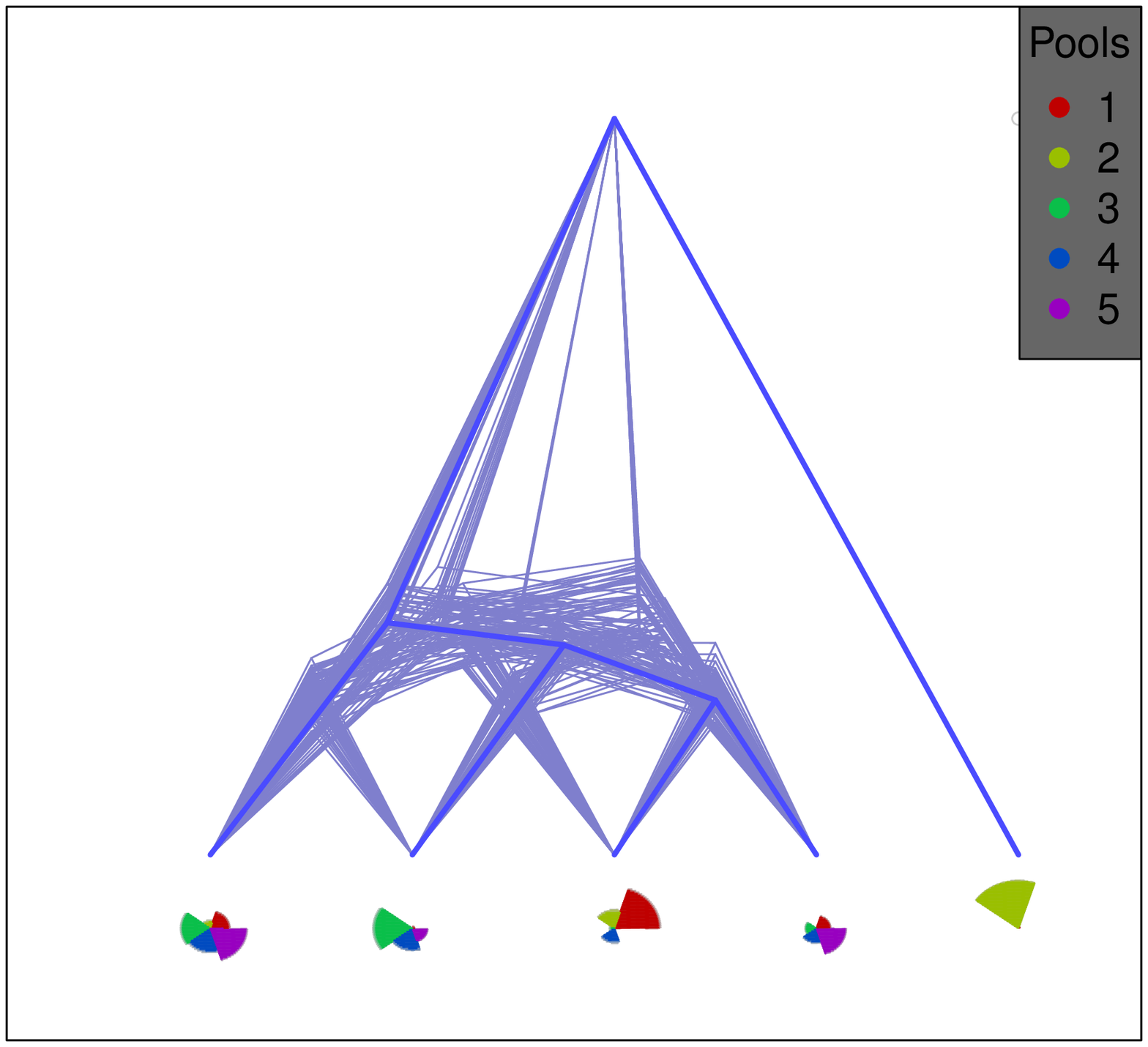}
	 }

	\caption{Location on simulated tree of SNPs for six sequence patterns (six above).  The branch width is proportional to number of SNPs.  The bottom figure shows the inferred lineage model.  }
	\label{island_ex}
\end{figure}

To provide a more in-depth understanding of the model's performance, we show a typical example for moderately high mixing data ($\psi=0.005$).  At the bottom of Figure \ref{island_ex}, we present the phylogenies and pool proportions inferred by the lineage model.  The simulation provides the branch where each SNPs relevant mutation occurred.  For each of the $2^5$ site patterns in the inferred model, we size the branches of the simulated tree by the number of SNPs with that inferred pattern.   We color branches with phylogenetic SNPs in red and with null SNPs in blue.  The lineages are numbered from left to right so that, for example, site pattern $(1,0,0,0,0)$ has SNP state $1$ for lineage $1$ and $0$ otherwise.  

We observe that the SNPs associated with a particular sequence pattern tend to fall on a single branch or a small number of proximate branches, indicating the model's preservation of topological structure.  The inferred model appears to recapitulate much of the relative location of these branches on the coalescent tree and also reflect appropriate pool proportions.  The null SNPs distribute relatively evenly over the tree's tips, except for one deep branch not captured by any sequence pattern.  We note that the inferred topology bimodal between two possible trees, likely driven by the locations of sequence patterns $(0,1,0,0,1)$ and $(0,0,0,1,1)$ both falling exclusively on a single branch within the coalescent tree.


\subsubsection{General performance}

Across the island coalescent simulations we find the performance of the model varies largely with the degree of mixing.  To ensure a uniform scale across simulations, we examine the average pairwise distance between SNPs with a common sequence pattern divided by the average pairwise distance over the entire tree.  We show the results in Supplementary Figure $5$.  When mixing is close to zero ($\psi = 0.0003$), the model reduces to a single sequence pattern per sample, phylogenetic sequence patterns strongly cluster on a single branch, and the inferred phylogenies show little topological uncertainty.  As $\psi$ increases the degree of localization decreases slightly for two orders of magnitude until rapidly increasing afterwards, with the model's topological uncertainty follows a similar progression.  For very high degrees of mixing, the localization for phylogenetic SNPs differs very little from that for null SNPs.  For all simulations, we find the null SNPs spread evenly over external or nearly-external branches.

\section{Empirical Examples}

\subsection{Green sulfur bacteria in an Antarctic lake}

The \emph{Chlorobium} genus comprises a class of of green sulfur bacteria that are one of the most photosynthetically productive microbial populations in anoxic aquatic environments.  We explore the composition of \emph{Chlorobium} strains from a set of metagenomic samples taken at differing depths within Ace lake, a pristine, anoxic, marine-derived, stratified lake in Antarctica formed approximately $5000$ years ago, as well as two nearby marine samples.  \cite{Lauro2011} provide a full description of the collection regime and an integrated, functional metagenomic analysis.  

We examine data from nine whole-genome sequence samples and their meta-data (443679.3-443687.3) downloaded from the MG-RAST server on October 15, 2011 \citep{Meyer2008}.  One freshwater sample contains no meta-data on sample depth collection.  For comparison against a \emph{Chlorobium} sequence, we downloaded the genome for \emph{Chlorobium limicola} from the NCBI Genome project website on October 20, 2011 \citep{Geer2010}.  We employ the \emph{de novo} variation detection algorithm Cortex to ascertain SNPs and their counts per sample.  We exclude four samples (4443679.3, 4443680.3, 4443681.3, 4443685.3) due to low coverage for most SNPs, leaving three lake samples and two marine samples.  We also remove indel variants and SNPs with fewer than $70$ read counts across the remaining samples, leaving $345$ SNPs for analysis.  

\begin{figure}
\centering
\includegraphics[scale = 0.8]{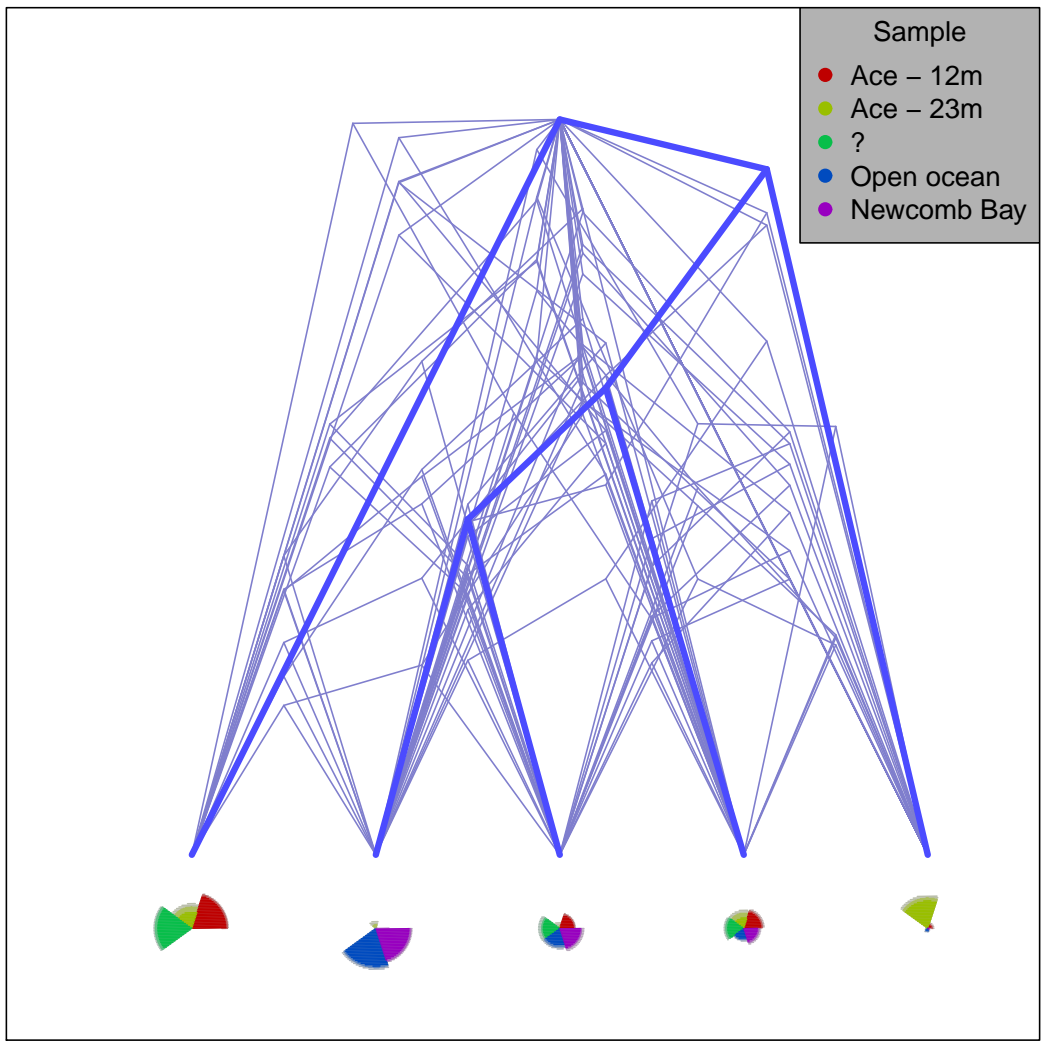}
\caption{Inferred lineage model for \emph{Chlorobium} data from Ace Lake and open ocean samples.}
\label{antarctic}
\end{figure}

Figure \ref{antarctic} shows the inferred lineage model for the five samples.  Lineage $1$ is found only in Ace Lake samples, while lineage $2$ is found only in marine samples.  We note that the deep divergence time of lineage $1$, substantially present within all lake samples, is consistent with long-term isolation of Ace Lake.  Lineage $5$ shows the presence of a unique strain within the $23$ m sample, consistent with previous analysis \citep{Lauro2011}.  Lineage $4$ appears to be present in all samples, although preferentially in those from the lake.  Lineage $3$ is similar, but has no contribution to the deep water sample.  We note that pool proportions of the unknown sample (green in the figure) indicate that it likely has a similar collection location to the $12$ m sample.  

\subsection{Mixed infections of \emph{Plasmodium falciparum} in northern Ghana}

\emph{Plasmodium falciparum} is the causative agent of most severe malaria world-wide and is endemic in large section of sub-Saharan Africa \citep{Snow2005}.   Examinations of infected blood samples frequently show multiple strains of parasites present within a single host, although the clinical import is debated \citep{Genton2008}.  A recent examination of whole-genome-sequenced parasite samples taken from clinical isolates indicates that the degree mixed infections varies strongly by geographic region, with western Africa exhibiting the highest values \citep{Manske2012}.  

\begin{figure}
\centering
\includegraphics[scale = 0.8]{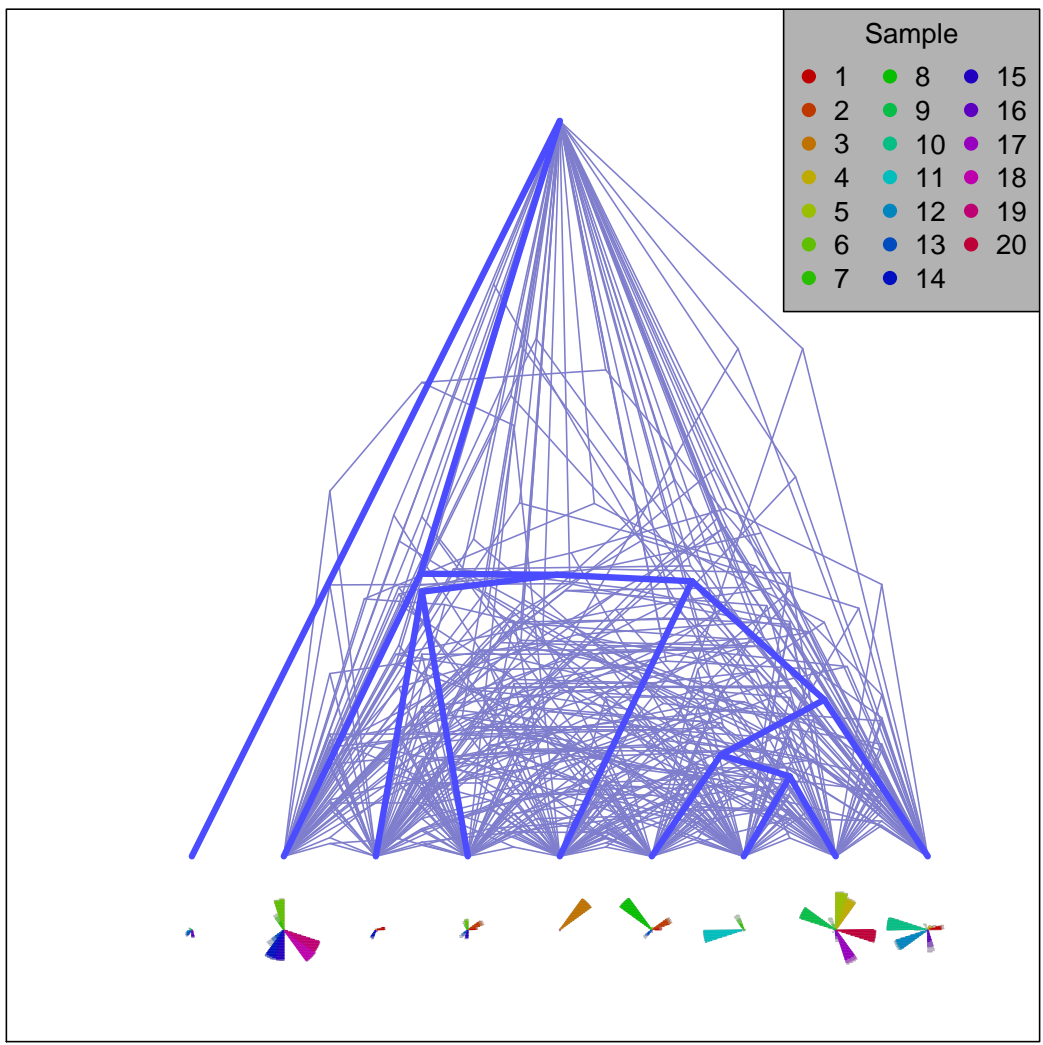}
\caption{Inferred lineage model for \emph{Plasmodium falciparum} apicoplast data from twenty clinical samples from northern Ghana.}
\label{ghana}
\end{figure}

Each \emph{Plasmodium falciparum} cell contains exactly two plastids: a mitochondrion and an apicoplast.  The apicoplast is a chloroplast-derived plastid necessary for essential heme metabolism.  Following methods in \citet{Manske2012}, we ascertain $123$ SNPs from the apicoplast within $20$ clinical isolates from the Kassena-Nankana district region of northern Ghana.  The model infers $9$ lineages shown in Figure \ref{ghana}.  Lineages $2$, $5$, and $8$ appear to by largely unmixed in their respective samples, while lineages $1$, $3$, $4$, and $9$ appear almost exclusive in mixed samples.  Lineages $6$ and $7$ appear in both mixed and unmixed samples.  We note that two lineages, $2$ and $8$, appear to dominate about half of the samples.  The topological uncertainty suggests that the data may not be yet sufficiently high quality for precise inference.  However, the output strongly indicates the presence of mixed infections, consistent with estimates from the nuclear genome, and suggests that the degree of mixture may vary with the underlying sequence.    

\subsection{\emph{Neisseria meningitidis} in sub-Saharan Africa}

We examine data from field samples of \emph{Neisseria meningitidis} collected on sequential visits to the Kassena-Nankana region in northern Ghana \citep{Leimkugel2007}.  \emph{N. meningitidis} exists as non-pathogenic flora in the naval cavity of about $10\%$ of adults \citep{Caugant1994}.  The same bacteria may exhibit hyper-virulent forms, leading to severe meningococcal meningitis\citep{Caugant2008}.  In sub-Saharan Africa, these virulent bacterial forms appear as an epidemic each $8-12$ years in the dry season, and researchers believe that these occurrences travel as ``waves'' across the continent from west to east \citep{Leimkugel2007}.  Researchers collected field samples from different individuals in two villages within KND from 1998 until 2008, although we examine only the subset of samples from 1998--2005. 

For sequencing, individual samples were pooled by villages and by years, giving us $10$ pools, with $2$ pools per epidemic season ($1998-1999$, $1999-2000$, \emph{et c.}).   Sequencing was performed on early Illumina technology and before the development of tags.  Using the read data, we ascertained SNPs using a novel \emph{de novo} assembly approach outlined in \citet{Ahiska2011}.  After cleaning for quality, we find $1099$ sites with a mean read count depth per site of $54.53$ reads.  Applying the lineage algorithm to this data yields the $5$ lineages shown in Supplementary Figure 6.  Pools $7-10$ correspond to years $2001-2002$, when researchers previously noted the advent of a new sequence type in KND.  The lineage model clearly separates the two epidemiological waves, as well as possible `subwaves' distinct from the dominant strains. 

\section{Discussion}
Biologists now produce enormous amounts of metagenomic data, investigating a range of systems from the microbiomes of beehives to the microflora of ocean vents.  Analyses of these data usually assess the proportions of living domains that read data can be uniquely mapped into, and compare across samples by contrasting their compositions.  These investigations naturally focus on macroevolution across species, phyla or families, where genomic change is so substantial amongst clades that each can be treated as fixed.  Often these studies focus only on the signal from a single gene, such as 16S sRNA.

In this paper, we consider metagenomics in the domain of microevolution, where genomic changes occur on the same time scale as environmental mixing, as in microbiomes, epidemics, or cancer cell lines.  This regime corresponds to the island coalescent model when the migration and mutation parameters are roughly on the same time scale.  We show that in this circumstance we can extract a meaningful phylogenetic signal.  The mixing rate is the key: for a small rate, the situation effectively reduces to a standard phylogenetics problem; when it is very high, we cannot parse out pool mixtures from the tree information; in between, we can make reliable inference.  However, we cannot yet provide precise guidelines about where this distinction occurs in biological systems, although we empirically observe that the model produces equal estimates of pool proportions across all lineages and high tree uncertainty when confronted with very low-quality or randomly generated data.  Our three empirical examples also give some guidance for appropriate applications of the model.   

In order to implement our model, we make a number of simplifying assumptions.  We assume that the pools are independent of each other, that SNPs are unlinked, and that recombination is non-existent.  In almost any biological experiment these postulates will be violated in some fashion.  However, all violations are not created equal.  Obligate recombination that occurs in sexual organisms will undoubtedly confound the model, rendering tree inference very questionable \citep{Schierup2000}.  The presence of moderate linkage among SNPs, on the other hand, will not prevent the model from functioning at all: we currently just neglect the additional information that would provide.  Similarly, some non-independence among pools will likely not harm the quality of inference under the model.    

We believe a place of possible improvement in our current implementation to be in our error model, where we treat every read as possessing possible sequencing errors.  While helpful in separating phylogenetic SNP variation from noise, we hope in the future to implement more biologically sophisticated models where low-frequency variants can be included. 
We conjecture that the inclusion of SNP count data where the states of multiple SNPs from a single organism, such as paired end data or longer read data, will help us fill this gap.  These reads provide strong evidence about the state of the lineages in reality and their inclusion into the model should permit better inference and more elaborate population models.  Our experience suggests that this extension will present a methodological challenge in the MCMC framework in finding approaches that efficiently mix over the parameter space.

Another natural extension is to weaken the assumption that the pools are independent.  In most studies we would expect \emph{a priori} that pools' composition will have strong correlations, induced by the sampling procedure in time or space or both.  Including these structures will provide strong indications about the pool composition, since nearby pools are presumably composed more similarly than distant ones.  We expect that a Gaussian Markov random field prior on the pool distribution determined by the graph representing the experimental sampling procedures (e.g. sampling times) will prove an efficient means of incorporating this information \citep{Rue2005}.  

\bibliographystyle{genetics}
\bibliography{biblio_lineage}
\clearpage
\section*{Tables}
\begin{table}[!ht]
\begin{tabular}{ll}
$\mathcal{D} = [d_{ijs}]$ & Data comprised of counts for each SNP $j$ within pool $i$ of type $s \in \{0,1\}$ \\
$i = 1, \cdots, N$ & Index and number of pools\\
$j = 1, \cdots, M$ & Index and number of SNPs\\
$k = 1, \cdots, K$ & Index and number of lineages \\
$\mathcal{L} = [l_{kj}]$ & Lineages composed by state of SNP $j$ in lineage $k$ \\
$\mathcal{T}$ & Phylogeny\\
$\tau$, $\{t_b\}$ & Topology and branch lengths for $\mathcal{T}$ \\
$T$ & The total branch length of $\mathcal{T}$ \\
$\lambda$ & Probability of a phylogenetic SNP \\
$\bar{\mathcal{D}}$, $\tilde{\mathcal{D}}$ & Phylogenetic and null SNP sets defined by $\mathcal{P}$ \\
$\mathcal{P}$ &Partition of SNPs into phylogenetic and null components\\
$\mathcal{S} = [s_{ik}]$ & Pool composition specified by pool proportion for pool $i$ and lineage $k$ \\
$p_{ij}$ & The uncorrected reference allele frequency for SNP $j$ in pool $i$ \\
$\eta$ & SNP error rate \\
$\tilde{p}_{ij}$ & The corrected reference allele frequency for SNP $j$ in pool $i$ \\
$\xi$ & Mutation rate \\
$\psi$ & Mixing rate in the island coalescent simulations
\end{tabular}
\caption{Symbols used in the model description.}
\label{table:symbols}
\end{table}

\begin{table}
\begin{center}
\begin{tabular}{ll}
Parameter & Values \\
\hline Number of SNPs ($M$)  & $25, \ 100, \ 250, \ 1000$ \\
Number of reads  & $2, \ 5, \ 10, \ 50$ \\
SNP error rate ($\eta$)  & $0, \ 0.001, \ 0.05, \  0.15$ \\
Mixture parameter ($\rho$)  &$0, \ 1.5, \ 4, \ 10$ \\
Fraction of null SNPs ($1-\lambda$) &  $0, \ 0.05, \ 0.15, \ 0.3$ \\
\end{tabular}
\end{center}
\caption{Parameter values used in model simulations}
\label{table:sim_paras}
\end{table}
\clearpage

\section*{Figures}

\clearpage

\clearpage

\clearpage

\end{document}